\documentclass[bimj,fleqn]{w-art}
\usepackage{times}
\usepackage{w-thm}
\usepackage[authoryear,round]{natbib}
\usepackage[colorlinks=true,linkcolor=black,citecolor=black,urlcolor=black,breaklinks=true]{hyperref}
\usepackage{breakurl}
\theoremstyle{plain}

\theoremstyle{definition}

\usepackage[]{graphicx}
\chardef\bslash=`\\ 

\begin{document}

\DOIsuffix{bimj.XXXXXXX}
\Volume{0}
\Issue{0}
\Year{2015}
\pagespan{1}{}

\keywords{Random-effects meta-analysis; Orphan disease; Bayesian statistics; Between-study heterogeneity; Coverage probability \\
\noindent \hspace*{-4pc} {\small\it }\\
\noindent\hspace*{-4.2pc} Supporting Information for this article is available from the author or on the WWW under\break \hspace*{-4pc} \underline{http://dx.doi.org/10.1022/bimj.XXXXXXX}
}  

\title[Meta-analysis of two studies\ldots]{Meta-analysis of two studies in the presence of heterogeneity\\ with applications in rare diseases}

\author[Friede {\textsl{et al.}}]{Tim Friede\inst{1}\footnote{Corresponding author: {\sf{e-mail: tim.friede@med.uni-goettingen.de}}, Phone: +49-551-39-4990, Fax: +49-551-39-4995}}
\author[]{Christian R\"{o}ver\inst{1}} 
\author[]{Simon Wandel\inst{2}}
\author[]{Beat Neuenschwander\inst{2}}
\address[\inst{1}]{University Medical Center G\"ottingen, Department of Medical Statistics, Humboldtallee~32, 37073~G\"ottingen, Germany}
\address[\inst{2}]{Novartis Pharma AG, Oncology, 4002~Basel, Switzerland}

\Receiveddate{xxx} \Reviseddate{yyy} \Accepteddate{zzz} 

\begin{abstract}
Random-effects meta-analyses are used to combine evidence of treatment effects from multiple studies. Since treatment effects may vary across trials due to differences in study characteristics, heterogeneity in treatment effects between studies must be accounted for to achieve valid inference. The standard model for random-effects meta-analysis assumes approximately normal effect estimates and a normal random-effects model. However, standard methods based on this model ignore the uncertainty in estimating the between-trial heterogeneity. In the special setting of only two studies and in the presence of heterogeneity we investigate here alternatives such as the Hartung-Knapp-Sidik-Jonkman method (HKSJ), the modified Knapp-Hartung method (mKH, a variation of the HKSJ method) and Bayesian random-effects meta-analyses with priors covering plausible heterogeneity values. The properties of these methods are assessed by applying them to five examples from various rare diseases and by a simulation study. Whereas the standard method based on normal quantiles has poor coverage, the HKSJ and mKH generally lead to very long, and therefore inconclusive, confidence intervals. The Bayesian intervals on the whole show satisfying properties and offer a reasonable compromise between these two extremes.
\end{abstract}

\maketitle                   






\section{Introduction}
  Meta-analyses are used to combine evidence of treatment effects from
  multiple studies. Since treatment effects may vary across trials due
  to some slight differences in study characteristics including study
  populations, trial designs, endpoints and standardization of
  treatments, heterogeneous treatment effects are quite natural and
  must be accounted for to achieve valid statistical
  inferences. Therefore, random-effects meta-analysis has become the
  standard to combine treatment effects from several studies when the
  presence of between-trial heterogeneity is suspected which is often
  the case.

  The standard model for random-effects meta-analysis assumes
  approximately normal effect estimates and a normal random-effects
  model, the \textsl{normal-normal hierarchical model}
  \citep{HedgesOlkin}. Based on this model, standard inference methods
  based on normal quantiles to construct confidence intervals for the
  combined effect ignore the uncertainty in the estimation of the
  between-study heterogeneity and they are only valid for large
  numbers of trials. However, the combination of only a few studies is
  quite common \citep{DaveyEtAl2011, TurnerEtAl2012}.  This is not
  only the case in rare diseases, but in this context it poses a
  particular challenge since increased levels of heterogeneity are
  common \citep{SimulationPaperDummyRef}.  For instance, in a recent
  systematic review by \citet{CrinsEtAl2014} six studies on acute
  graft rejections and three studies on steroid-restistant rejections
  were combined in random-effects meta-analyses to assess the efficacy
  and safety of Interleukin-2 receptor antibodies for
  immunosuppression following liver transplantation in children. All
  studies were controlled, but only two were randomised as it is often
  the case in paediatrics. Furthermore, there were some differences
  between the studies with respect to their control groups and other
  design characteristics suggesting some degree of between-trial
  heterogeneity.

  For random-effects meta-analyses with few studies methods based on
  $t$-distributions have been suggested \citep{FollmannProschan1999,
    HartungKnapp2001a, HartungKnapp2001b, KnappHartung2003,
    SidikJonkman2003}.
  Furthermore, the use of priors covering plausible between-trial
  standard deviations has been advocated when dealing with few studies
  \citep{SpiegelhalterEtAl,SimulationPaperDummyRef,NeuenschwanderEtAl2010,SchmidliEtAl2014}.

  Here we consider the special case of only two studies which has
  recently attracted some attention
  \citep{GonnnermannEtAl2015}. Examples for meta-analyses of two
  studies include the summary of two pivotal studies of a clinical
  development programme \citep{EMEA1998,EMEA2001,ICH2002}. As we will
  see when discussing several examples below, meta-analyses of two
  studies are not uncommon in orphan diseases. For instance, two
  randomised controlled trials were included in the systematic review
  by \citet{CrinsEtAl2014} and the Cochrane Review by
  \citet{MillerEtAl2012} on Riluzole in amyotrophic lateral sclerosis
  (ALS). In the presence of heterogeneity, however, the meta-analysis
  of only two studies may be considered an unsolved problem
  \citep{GonnnermannEtAl2015}. Therefore, we assess here the
  performance of alternative approaches to real-life examples from
  rare diseases and by exploring their charactersistics in an
  extensive simulation study. Based on these findings we give some
  recommendations on how to approach the problem successfully in
  practice.

  The paper is organised as follows. In Section~\ref{sec:methods} the
  statistical model is introduced and methods for frequentist and
  Bayesian inference are reviewed. Five examples in various rare
  diseases are presented in Section~\ref{sec:examples} before an
  extensive simulation study is presented in
  Section~\ref{sec:sims}. In Section~\ref{sec:discussion} we close
  with a brief discussion of the findings.

\section{Methodology} \label{sec:methods}
\subsection{Notation and statistical model}
  Standard meta-analytic models assume either a common (fixed) effect
  or random effects across studies. For the latter, the normal-normal
  hierarchical model (NNHM) is the most popular. At the first level,
  the sampling model assumes approximately normally distributed
  estimates $Y_1,\ldots,Y_k$ for the trial-specific parameters
  $\theta_1,\ldots,\theta_k$
  \begin{equation}
    Y_j \vert \theta_j \;\sim\; \mathrm{N}(\theta_j,s_j^2), \quad j=1,\ldots,k .
  \end{equation}
  Here, we will follow the standard assumption which treats the
  standard errors~$s_j$ as known, although this could be relaxed if
  necessary.  At the second level, the parameter model assumes
  normally distributed study effects
  \begin{equation}
    \theta_j \vert \mu, \tau \;\sim\; \mathrm{N}(\mu,\tau^2),  \quad j=1,\ldots,k.
  \end{equation}
  The between-trial standard deviation $\tau$ determines the degree of
  heterogeneity across studies.  If the parameter of interest is $\mu$
  (rather than the study effects $\theta_j$), inference can be
  simplified by using the marginal model
  \begin{equation}
    Y_j \vert \mu,\tau \;\sim\; \mathrm{N}(\mu,s_j^2+\tau^2), \quad j=1,\ldots,k .
  \end{equation}
  The two main approaches to infer $\mu$ and the nuisance parameter
  $\tau$ are frequentist and Bayesian. If $\tau$ were known,
  frequentist and Bayesian (with a non-informative prior for $\mu$)
  conclusions would be analogous. In fact, in the frequentist setting
  \begin{equation} \label{eqn:est.mu}
    \hat{\mu} = \sum_{j=1}^k w_jY_j \bigg/ \sum_{j=1}^k w_j \;\sim\;
    \mathrm{N}(\mu,1/w_{+}),
    \qquad w_j = 1 \big/ \bigl(s_j^2+\tau^2\bigr), \qquad j=1,\ldots k,
  \end{equation}
  where $w_j$ are inverse-variance (precision) weights, and $w_+ =
  \sum_{j=1}^k w_j$ is the total precision; the respective variance
  $1/w_+$ is important to construct confidence intervals for $\mu$, as
  shown in Section~\ref{methods:freq}.  The Bayesian result (posterior
  distribution) is
  \begin{equation}
    \mu \vert Y_1,\ldots,Y_k \;\sim\; \mathrm{N}\Biggl(\sum_{j=1}^k w_jY_j \bigg/ \sum_{j=1}^k w_j,\;1/w_{+}\Biggr) .
  \end{equation}
  For unknown $\tau$, this frequentist-Bayesian ``equivalence'' breaks
  down, since the two approaches handle estimation uncertainty for
  $\tau$ differently.

\subsection{Frequentist inference} \label{methods:freq}
  For unknown $\tau$ we first consider frequentist methods to infer
  $\mu$, which comprise two steps.
  \begin{enumerate}
    \item[(1)] An estimate $\hat{\tau}$ is derived, from which
      estimated weights $\hat{w}_j = 1/(s_j^2+\hat{\tau}^2)$ and a
      corresponding estimate $\hat{\mu}$ in~(\ref{eqn:est.mu}) are
      obtained. Various estimators for $\tau$ have been proposed (for
      an overview see
      \citet{DerSimonianKacker2007,Rukhin2012,VeronikiEtAl2015}), the
      most prominent being the moment-estimator due to DerSimonian and
      Laird (DL).  Alternatives are the maximum likelihood (ML)
      estimator, the restricted maximum-likelihood estimator (REML),
      and the Paule-Mandel estimator (PM). While these estimates can
      differ considerably, for the special case of two trials they
      coincide \citep{Rukhin2012}.  We will refer to this common
      estimate
      \begin{equation} 
        \hat{\tau}^2 = \frac{(y_1-y_2)^2-s_1^2-s_2^2}{2}. 
      \end{equation}
      as the DL estimate, whereby negative values are set to zero.
    \item[(2)] A confidence interval for $\mu$ is then derived. Here
      we will investigate three methods.
      \begin{enumerate}
        \item[(i)] The simplest approach, which was proposed in the
          seminal paper by \citet{DerSimonianLaird1986}, uses the
          following normal approximation
          \begin{equation}
            \mbox{(DL)} \qquad \hat{\mu} \pm \hat{\sigma}_{\mu} \, z_{(1-\alpha/2)}, \qquad
                        \mbox{where} \quad \hat{\sigma}_{\mu}^2 = 1 \Big/ \sum_{j=1}^k \hat{w}_j, 
          \end{equation}
          and $z_p$ is the $p$-quantile of the standard normal
          distribution.  This method is known to be problematic for
          small~$k$, since it ignores the uncertainty of $\hat{\tau}$
          and will therefore give too narrow confidence intervals and
          inflated type-I errors.
        \item[(ii)] Various improvements using a $t$-distribution with
          $k\!-\!1$ degrees of freedom and alternative estimators
          for~$\sigma_{\mu}$ have been proposed
          \citep{HartungKnapp2001a,HartungKnapp2001b,KnappHartung2003,SidikJonkman2003}.
          The HKSJ confidence interval is given by
          \begin{equation} \label{eqn:HKSJ}
            \mbox{(HKSJ)} \qquad \hat{\mu} \pm \tilde{\sigma}_{\mu} \, t_{k-1,(1-\alpha/2)}, \qquad
            \mbox{where} \quad \tilde{\sigma}_{\mu}^2 = \frac{1}{k-1}\sum_{j=1}^k \hat{w}_j
            (y_j-\hat{\mu})^2 \bigg/ \sum_{j=1}^k \hat{w}_j,
          \end{equation}
          and $t_{k-1,(1-\alpha/2)}$ is the
          $(1\!-\!\alpha/2)$-quantile of the Student\mbox{-}$t$
          distribution with $k\!-\!1$ degrees of freedom.  It works
          well for any number of studies if study-specific standard
          errors $s_j$ are of similar magnitude. Otherwise, coverage
          probabilities can be below the nominal level.
\item[(iii)]
  To address the limitations of the HKSJ method, a modified interval 
  \begin{equation}
    \mbox{(mKH)} \qquad\hat{\mu} \pm \sigma_{\mu}^{\star} \, t_{k-1,(1-\alpha/2)}, \qquad
     \mbox{where} \quad
    \sigma_{\mu}^{\star} = \mbox{max}\{\hat{\sigma}_{\mu},\tilde{\sigma}_{\mu}\}
  \end{equation}
  has been proposed \citep{RoeverKnappFriede2015}.  By taking
  the maximum of $\hat{\sigma}_{\mu}$ and $\tilde{\sigma}_{\mu}$, the
  problems of undercoverage and occasional counterintuitive results can be resolved.
\end{enumerate}
\end{enumerate}

\subsection{Bayesian inference} \label{methods:bayes}
  In the Bayesian framework, uncertainty of $\tau$ is automatically
  accounted for. Inference for $\mu$ and $\tau$ is captured by the
  joint posterior distribution of the two parameters, from which the
  marginal distribution of $\mu$ is used to derive, for example, point
  estimates and probability intervals for $\mu$. While automatic, the
  approach requires sensible prior distributions for $\mu$ and
  $\tau$. For the main parameter $\mu$, we will use a noninformative
  (improper) uniform prior.

  For $\tau$, however, the choice of prior is critical, in particular
  if the number of studies is small
  \citep{DiasEtAl2012,DiasEtAl2014,TurnerEtAl2015}. For the case of
  two studies and in the absence of relevant external data,
  information about between-trial heterogeneity is clearly very
  small. Therefore, the main feature of the Bayesian approach is its
  ability to average over the uncertain between-trial
  heterogeneity. This requires a prior distribution for $\tau$ that
  covers plausible between-trial standard deviations. If information
  about heterogeneity is weak, the 95\% prior interval should capture
  small to large heterogeneity.

  \begin{table} [ht]
    \begin{center}
      \caption{\label{tab:HN}Characteristics of the two half-normal priors for log-odds-ratios.}
      \begin{tabular}{ccc}
        \hline \hline
        prior   & median & 95\%-interval \\ \hline
        HN(0.5) & 0.337  & (0.016, 1.12) \\
        HN(1.0) & 0.674  & (0.031, 2.24) \\ \hline \hline
      \end{tabular}
    \end{center}
  \end{table}

  What constitues small to large heterogeneity depends on the
  parameter scale. For example, for log-odds-ratios (see examples in
  Section 3), values for $\tau$ equal to 0.25, 0.5, 1, and 2 represent
  moderate, substantial, large, and very large heterogeneity. We will
  use two half-normal (HN) prior distributions
  \citep{SpiegelhalterEtAl} in the examples (Section 3) and the
  simulation study (Section 4), with scale parameters 0.5 and 1.0; for
  prior medians and 95\%-intervals see Table~\ref{tab:HN}.
  The HN(0.5) prior captures heterogeneity values typically seen in
  meta-analyses of heterogeneous studies, and will therefore be a
  sensible choice in many applications. If very large between-trial
  heterogeneity is deemed possible, the more conservative HN(1.0)
  prior may be advised.

\section{Applications in rare diseases} \label{sec:examples}
\subsection{Introductory remarks}
  In this section we discuss five real-life examples of meta-analyses
  of two randomized controlled trials in various rare conditions. The
  first two are from the literature whereas the other three examples
  are based on US Food and Drug Administration (FDA) approvals in
  orphan diseases for the following drugs: Romiplostim, Mozobil, and
  Krystexxa.  In neither of these approvals, a formal meta-analysis
  was presented in the official documents.

  All examples have a binary endpoint comparing a treatment (T) to a
  control (C).  The following normal approximation on the
  log-odds-ratio scale
  \begin{equation}
    Y = \log \biggl(\frac{r_\mathrm{T}(n_\mathrm{C}-r_\mathrm{C})}{r_\mathrm{C}(n_\mathrm{T}-r_\mathrm{T})} \biggl),
    \qquad
    s^2 = \frac{1}{r_\mathrm{T}} + \frac{1}{n_\mathrm{T}-r_\mathrm{T}} + \frac{1}{r_\mathrm{C}} + \frac{1}{n_\mathrm{C}-r_\mathrm{C}}
  \end{equation}
  will be used, where $r$ and $n$ denote the number of responders and
  number of subjects, respectively.
  \begin{figure}[ht]
    \begin{center}
      \includegraphics[width=0.49\linewidth]{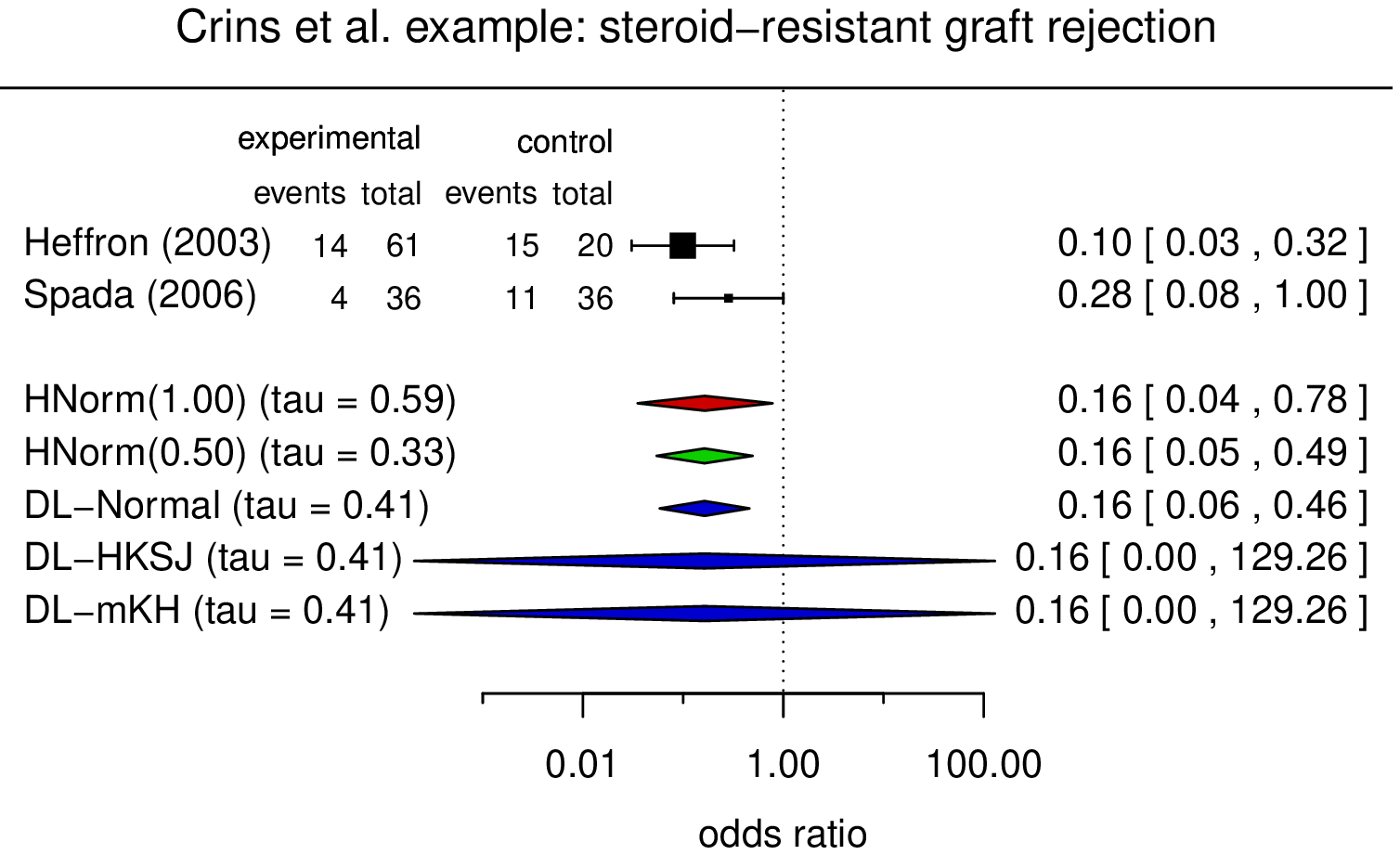} 
      \includegraphics[width=0.49\linewidth]{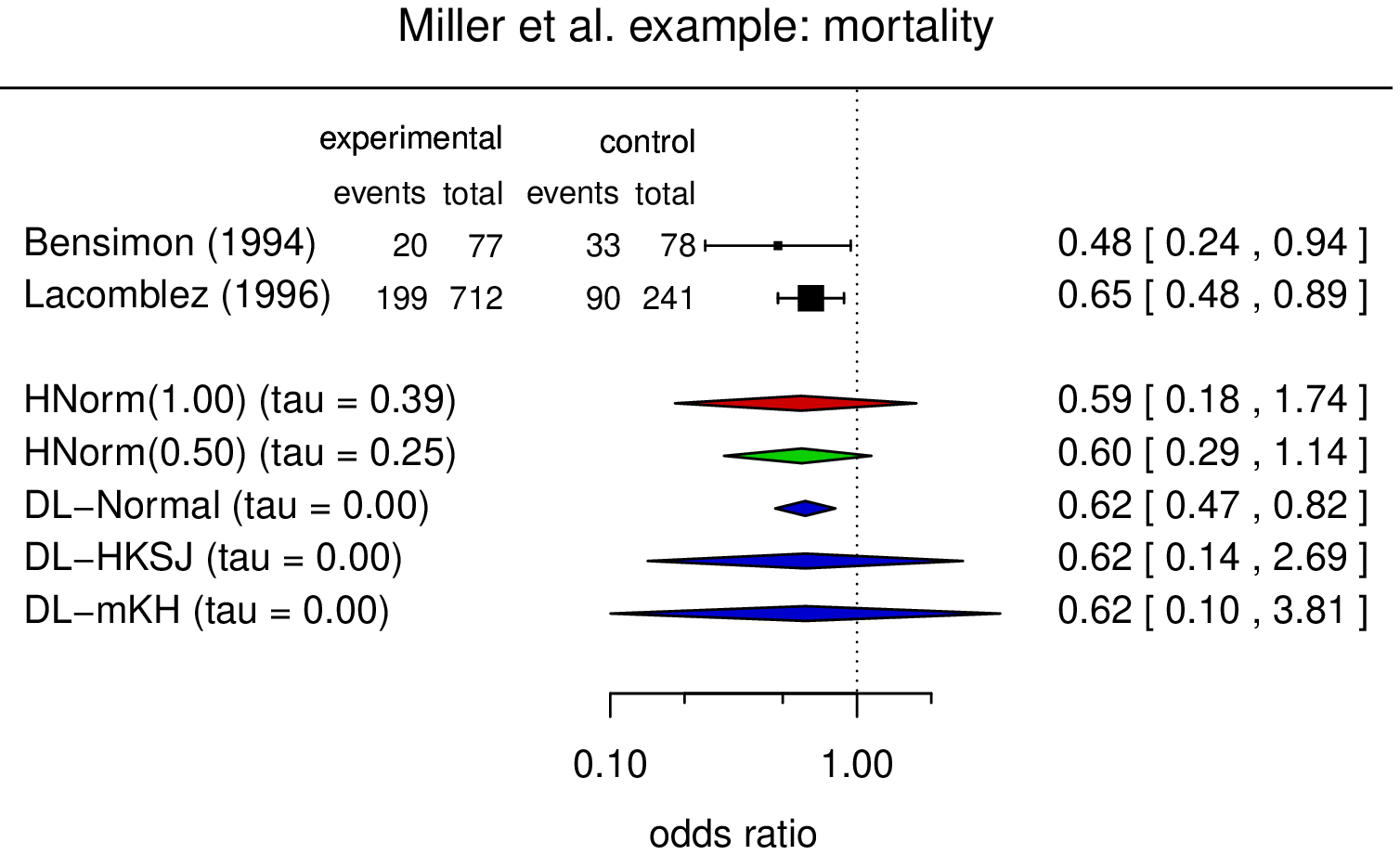} 
      \includegraphics[width=0.49\linewidth]{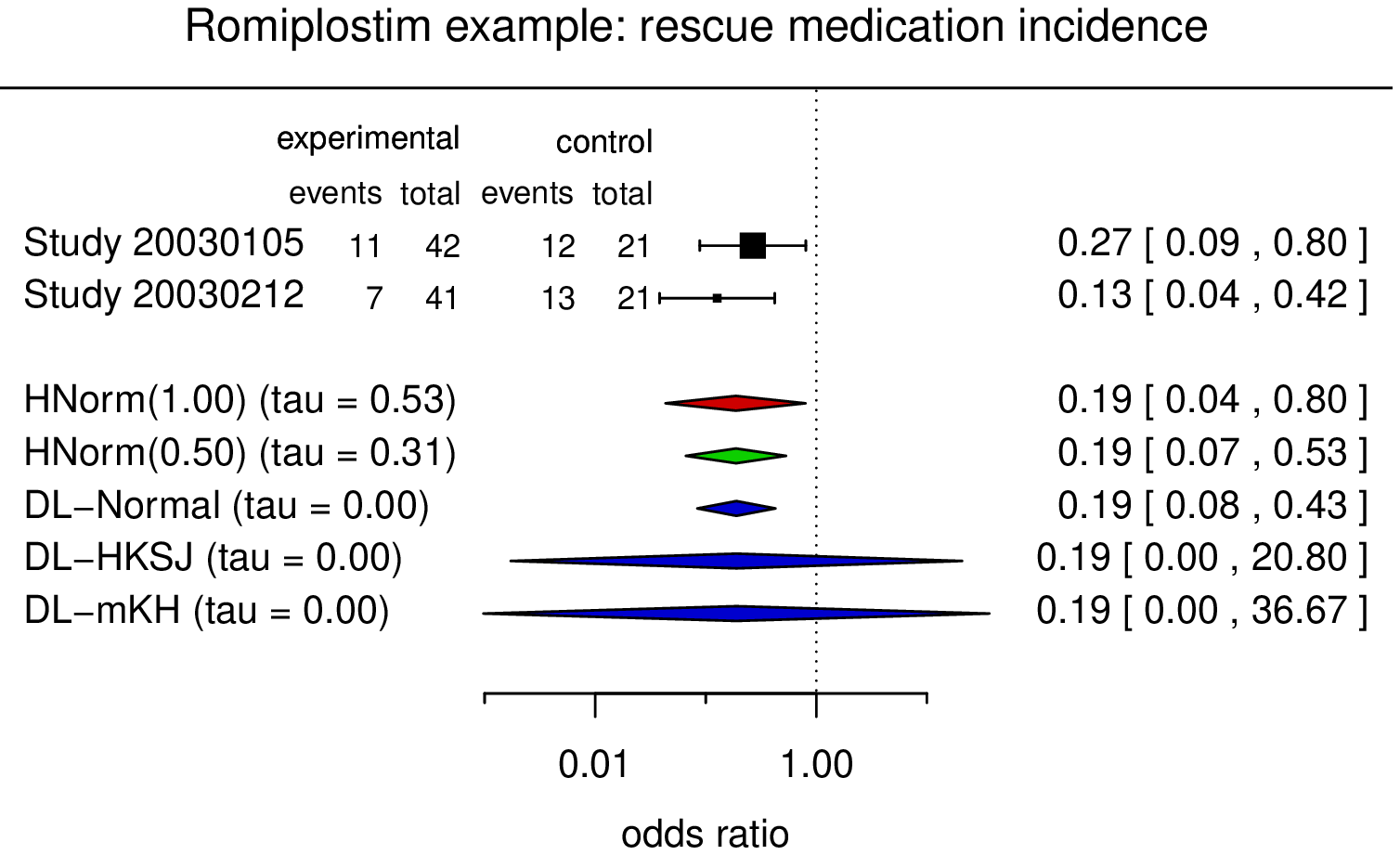} 
      \includegraphics[width=0.49\linewidth]{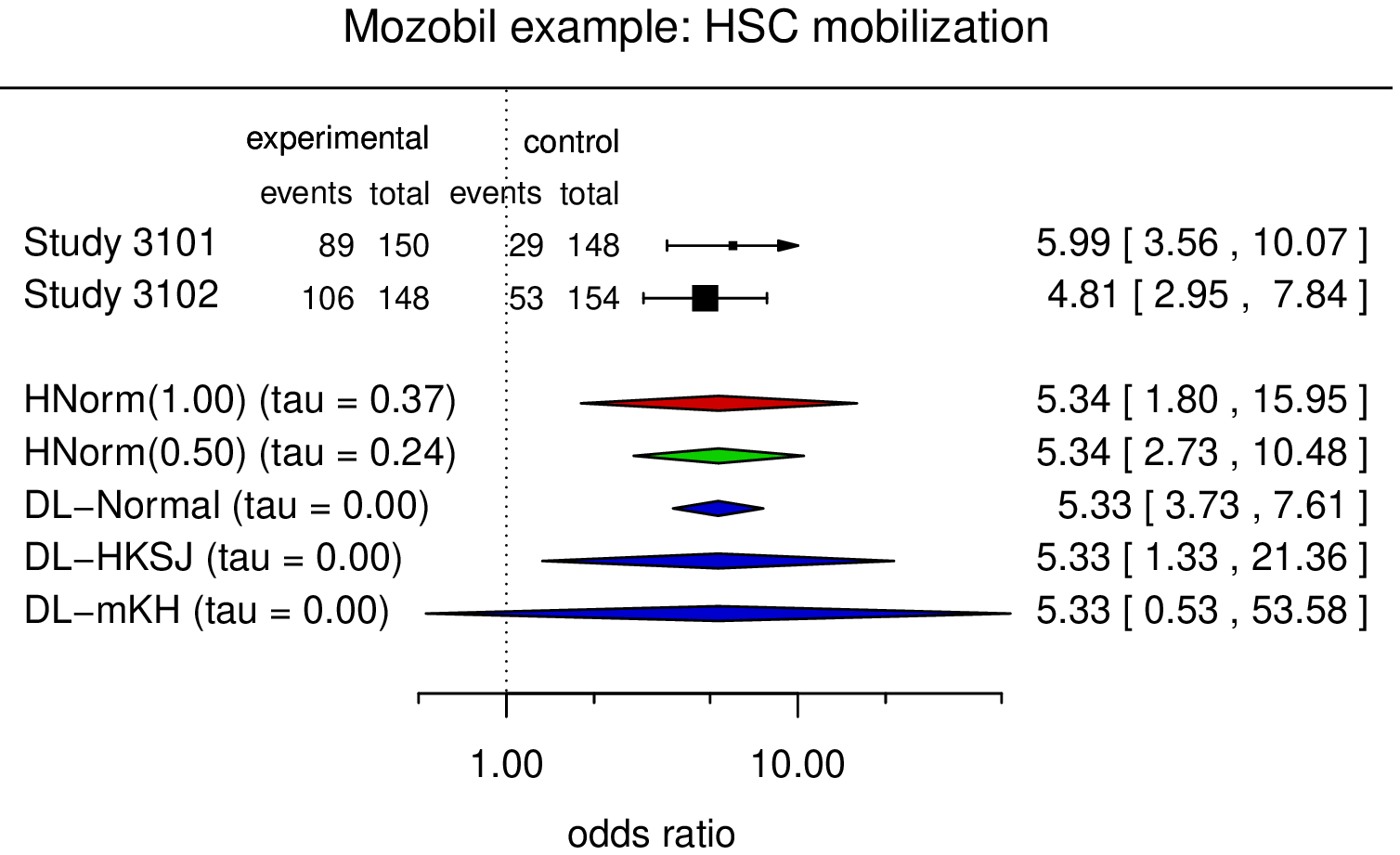} 
      \includegraphics[width=0.49\linewidth]{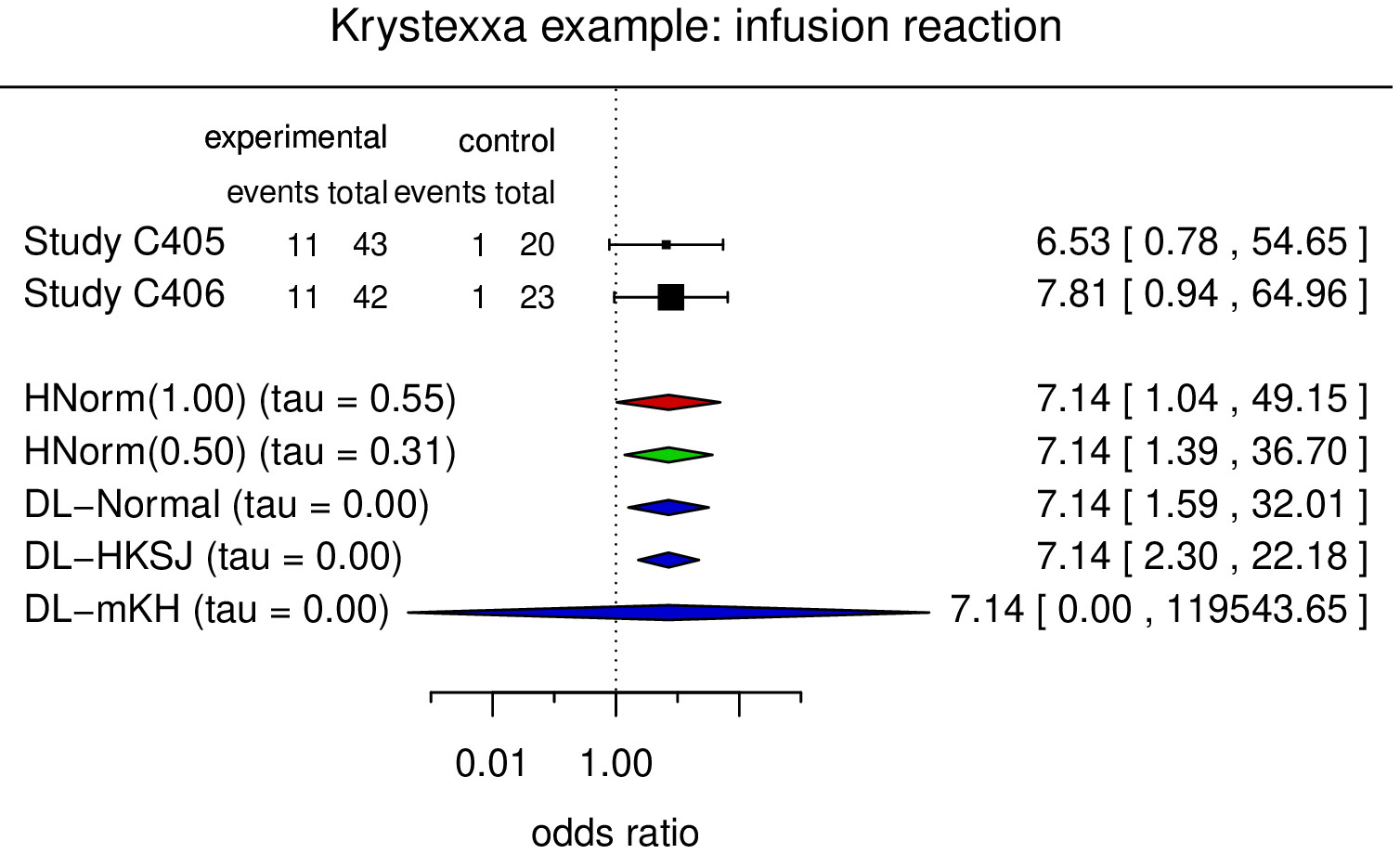} 
      \hspace{0.49\linewidth}
      \caption{\label{fig:examples} Forest plots for the five examples
        from rare diseases with various combined estimates of the
        treatment effect. The top two rows show the underlying data
        (numbers of cases and events in experimental and control
        groups) and illustrate the resulting estimates with their 95\%
        confidence intervals. The following rows show the different
        combined estimates along with the estimated amount of
        heterogeneity (posterior medians for the Bayesian
        approaches).}
    \end{center}
  \end{figure}

\subsection{Systematic review of Interleukin-2 receptor antibodies in pediatric liver transplantation \citep{CrinsEtAl2014}}
  \citet{CrinsEtAl2014} conducted a systematic review of controlled
  trials providing evidence on the efficacy and safety of
  immunosuppressive therapy with Interleukin-2 receptor antibodies
  (IL-2RA) Basiliximab and Daclizumab following liver transplantation
  in children. Six studies were included in a meta-analysis of acute
  graft rejections, of which only two were randomized (Heffron 2003,
  Spada 2006). In both studies about 80~patients were randomized, with
  2:1 allocation in \citet{HeffronEtAl2003} and 1:1 allocation in
  \citet{SpadaEtAl2006}. For the purpose of illustration we present
  here a meta-analysis of the two randomised studies in
  Figure~\ref{fig:examples}.

  Both studies yielded statistically significant results. However,
  there were some differences in the estimated odds ratios resulting
  in moderate to substantial estimates of the between-trial standard
  deviation. Although the two studies were statistically significant,
  the HKSJ and mKH methods result in confidence intervals that include
  the null hypothesis and are extremely wide (0--129 on the odds ratio
  scale). In contrast, the other three meta-analyses yield
  statistically significant results with the standard method based on
  normal quantiles giving the shortest confidence interval.

\subsection{Cochrane review of Riluzole in ALS \citep{MillerEtAl2012}}
  A Cochrane Review of Riluzole for amyotrophic lateral sclerosis
  (ALS) combined two randomized, placebo controlled, double-blind
  trials \citep{BensimonEtAl1994, LacomblezEtAl1996} with information
  on 12-month mortality in a meta-analysis
  \citep{MillerEtAl2012}. \citeauthor{MillerEtAl2012} combined the
  three active doses of the dose-ranging study by
  \citet{LacomblezEtAl1996} into one group for the purpose of the
  presented analysis. Whereas they used relative risks for their
  analyses we present here the results in terms of odds ratios
  (Figure~\ref{fig:examples}).

  As with the previous example both studies demonstrated statistically
  significant effects of the experimental drug over control (see
  Figure~\ref{fig:examples}). Whereas in the previous example the DL
  estimate of the between-trial heterogeneity was positive, here it is
  zero.  In comparison to the \citet{CrinsEtAl2014} example, here the
  HKSJ and the mKH methods are more informative as they are not quite
  as long. However, they are still considerably longer than the
  Bayesian intervals, which appear to be conservative since they
  include odds ratios of 1 although the confidence intervals of the
  individual studies both exclude 1.

\subsection{FDA approval in orphan disease: Romiplostim}
  Romiplostim \citep{FDA-romiplostim} was approved to treat Idiopathic
  Thrombocytopenic Purpura based on two 2:1 randomized studies. The
  two studies, 20030105 and 20030212, enrolled splenectomized and
  non-splenectomized patients, but were similar in their designs.

  Here we focus on patients requiring rescue medications (a secondary
  endpoint). Both studies showed statistically significant odds ratios
  (ORs): 0.27 (0.09, 0.80) for 20030105 and 0.13 (0.04, 0.42) for
  20030212 (Figure~\ref{fig:examples}). The ratio of ORs is~2.12,
  suggesting that between-trial heterogeneity should be considered.
  However, the frequentist estimate $\hat{\tau}$ is zero, resulting in
  a narrow confidence interval for $\mu$. On the other hand, the HSKJ
  and mKN intervals are very wide and do not allow sensible
  conclusions about the treatment effect.  The respective Bayesian
  intervals are much more plausible. Additionally, for both Bayesian
  analyses, the posterior medians for $\tau$ are smaller than the
  respective prior medians, indicating that the two half-normal priors
  do not unduly favour small homogeneity.

\subsection{FDA approval in orphan disease: Mozobil}
  Mozobil \citep{FDA-mozobil} was approved for the mobilization of
  hematopoietic stem cells in patients with lymphoma and multiple
  myeloma.  The two 1:1 randomized studies were conducted in two
  different indications: 3101 in Non-Hodgkin's Lymphoma, and 3102 in
  Multiple Myeloma. However, no differential treatment effect with
  respect to the primary endpoint was expected, which justifies a
  meta-analysis of the two studies.

  Both studies show statistically significant odds ratios
  (Figure~\ref{fig:examples}).  The ratio of the odds ratios is 1.25,
  suggesting possibly small between-trial heterogeneity.
  Unsurprisingly, the frequentist estimate $\hat{\tau}$ is zero. The
  HKSJ and mKH methods again provide very wide (but fairly different)
  CIs: the HKSJ method leads to a conclusive result, whereas the more
  conservative mKH does not; this is clearly implausible, since both
  studies showed highly significant results.  The respective Bayesian
  intervals are much narrower, suggesting a sensible compromise
  between the rather extreme (narrow and wide) frequentist
  counterparts.

\subsection{FDA approval in orphan disease: Krystexxa}
  For Krystexxa \citep{FDA-krystexxa}, two 2:2:1 randomized studies
  were used for approval. Here, we consider only one of two treatment
  arms (approved dose of 8mg every 2 weeks) and analyze a safety
  endpoint (infusion reaction). The two studies showed the following
  ORs: 6.55 (0.78, 54.60) for C405 and 7.77 (0.94, 64.72) for C406
  (Figure~\ref{fig:examples}), which suggest an increase in infusion
  reaction.

  In this example, the HKSJ and mKH intervals, which are usually very
  wide, give completely different answers. The HKSJ interval is even
  narrower than the interval based on normal approximations, whereas
  the mKH interval is unrealistically wide. The overly narrow HKSJ
  interval is due to the similar log-odds-ratios $y_j$ (1.88 and
  2.05), which lead to a very small estimate
  $\tilde{\sigma}_{\mu}=0.089$ in equation~(\ref{eqn:HKSJ}); the
  classical estimate $\hat{\sigma}_{\mu}=0.765$, which is used for
  mKH, is dramatically larger.

\subsection{Some concluding remarks on the examples}
  In this section we presented five examples from a range of rare
  diseases. In each of these, two studies were combinded in
  meta-analyses in situations where between-study heterogeneity had to
  be suspected to be present. Still the DL estimator for the
  between-study heterogeneity was zero in four out of the five
  examples.  Furthermore, the standard approach based on normal
  quantiles led to the shortest intervals in all but the Krystexxa
  example, in which the HKSJ interval was very narrow.  Otherwise the
  HSKJ and mKH methods yielded overall long to extremely long
  confidence intervals not conveying useful information on the size of
  the treatment effect. This is not surprising, since the $97.5 \%$
  quantile of a $t$-distribution with 1~degree of freedom is
  about~12.7. Although the Bayesian intervals appeared to be
  conservative, they led to interpretable results and a sensible
  compromise between the very short intervals based on normal
  quantiles and the often extremly long intervals based on
  $t$-quantiles.

\section{Simulation study} \label{sec:sims}
\subsection{Setup}
  For the simulation study of this section, we used the NNHM of
  Section~\ref{sec:methods}.  Simulations were limited to the case of
  two studies. The study sample sizes $n_1$ and $n_2$ were set to 25,
  100, or 400, which leads to six different combinations of
  $(n_1,n_2)$.  In the following figures (2--4), the first rows show
  results for equally sized studies, while the second rows illustrate
  the imbalanced settings.  Standard errors for the estimated
  log-odds-ratios $Y_j$ were set to $2/\sqrt{n_1}$ and
  $2/\sqrt{n_2}$. Without loss of generality, $\mu$ was set to zero.
  In terms of the ``relative'' amount of heterogeneity~$I^2$
  \citep{HigginsThompson2002}, the different settings correspond to
  $I^2 \in [0.20, 0.80]$ for $\tau=0.2$, and to $I^2 \in [0.86,0.99]$
  for $\tau=1.0$.  The number of simulations, which were performed
  using~\textsf{R}, was $15\,000$. Simulation results are shown for
  the bias of $\tau$ estimates, the fraction of $\tau$ estimates equal to
  zero, the coverage probabilities and the interval lengths for~$\mu$.

\subsection{Bias in estimators of the between-study heterogeneity $\tau$}
\begin{figure}[ht]
\begin{center}
\includegraphics[width=0.3\linewidth]{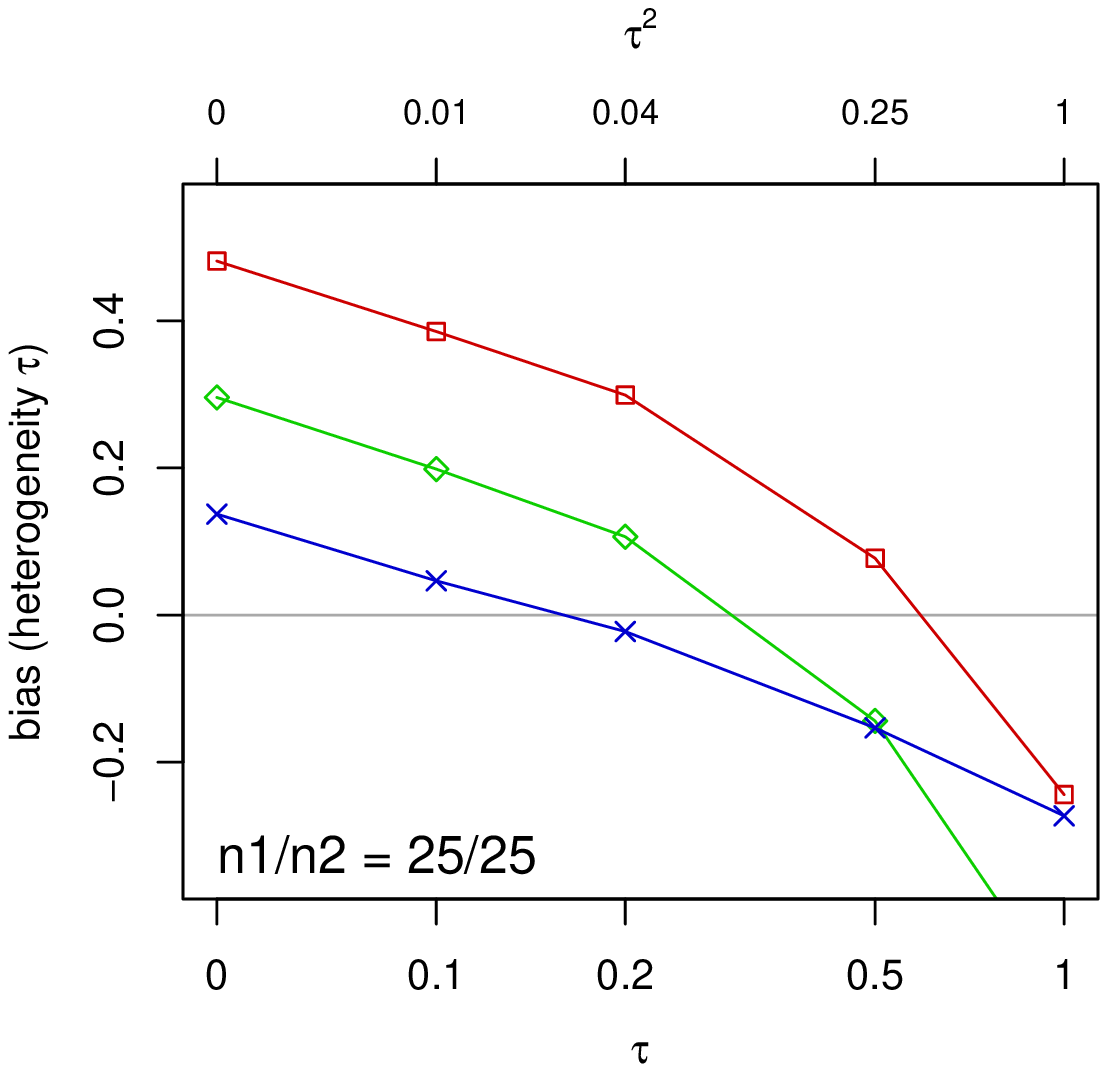} 
\includegraphics[width=0.3\linewidth]{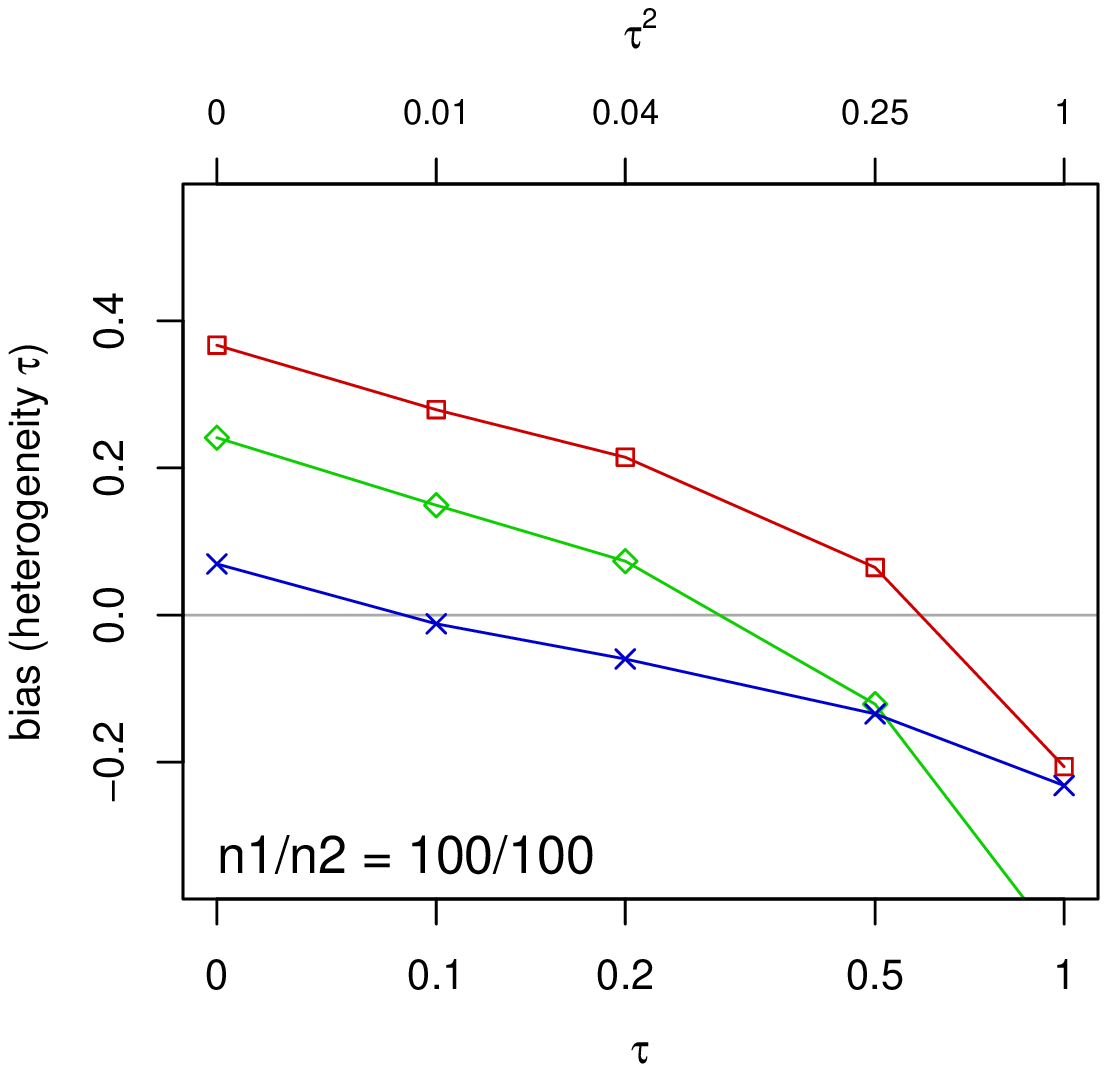} 
\includegraphics[width=0.3\linewidth]{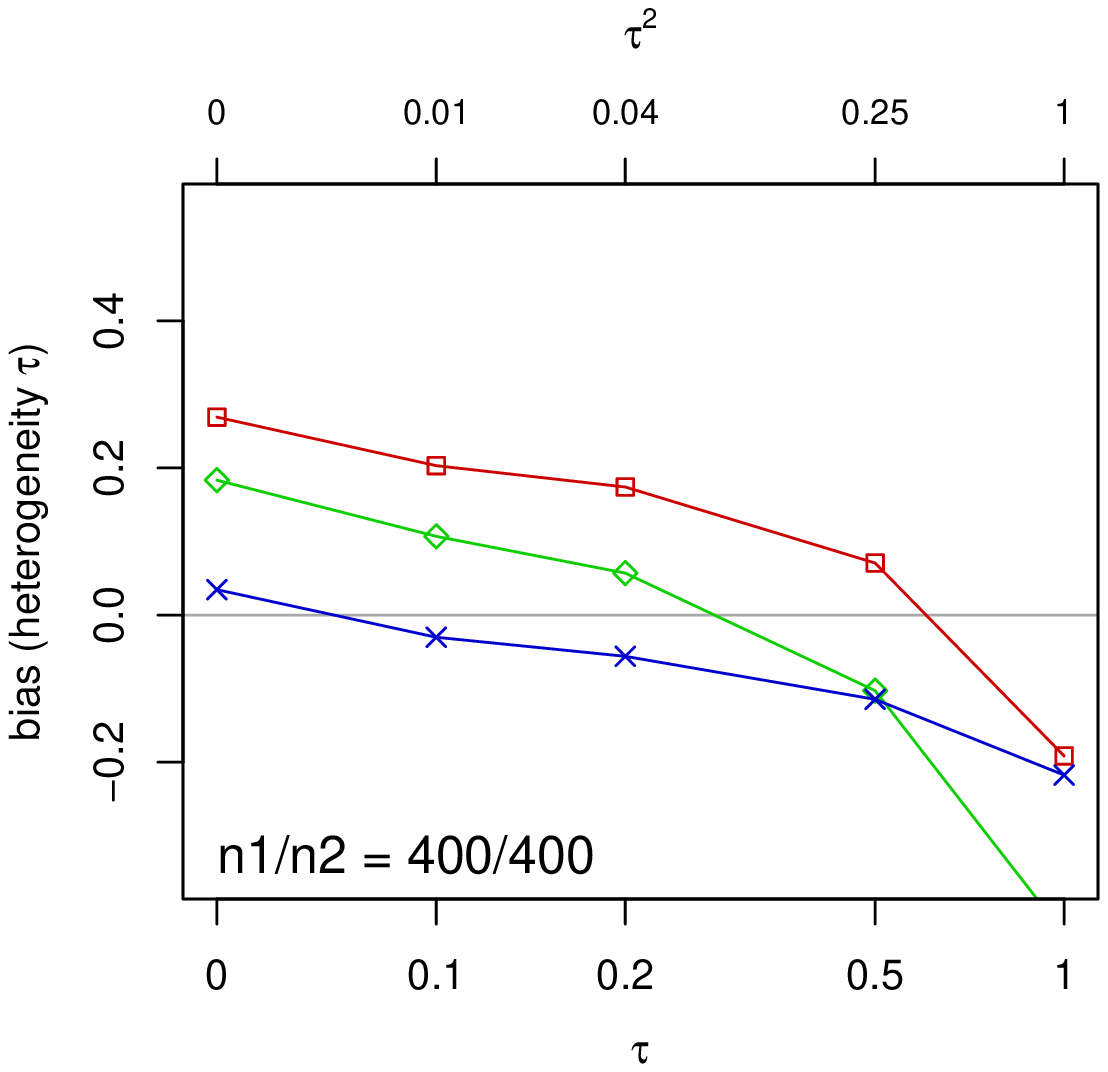} \\
\includegraphics[width=0.3\linewidth]{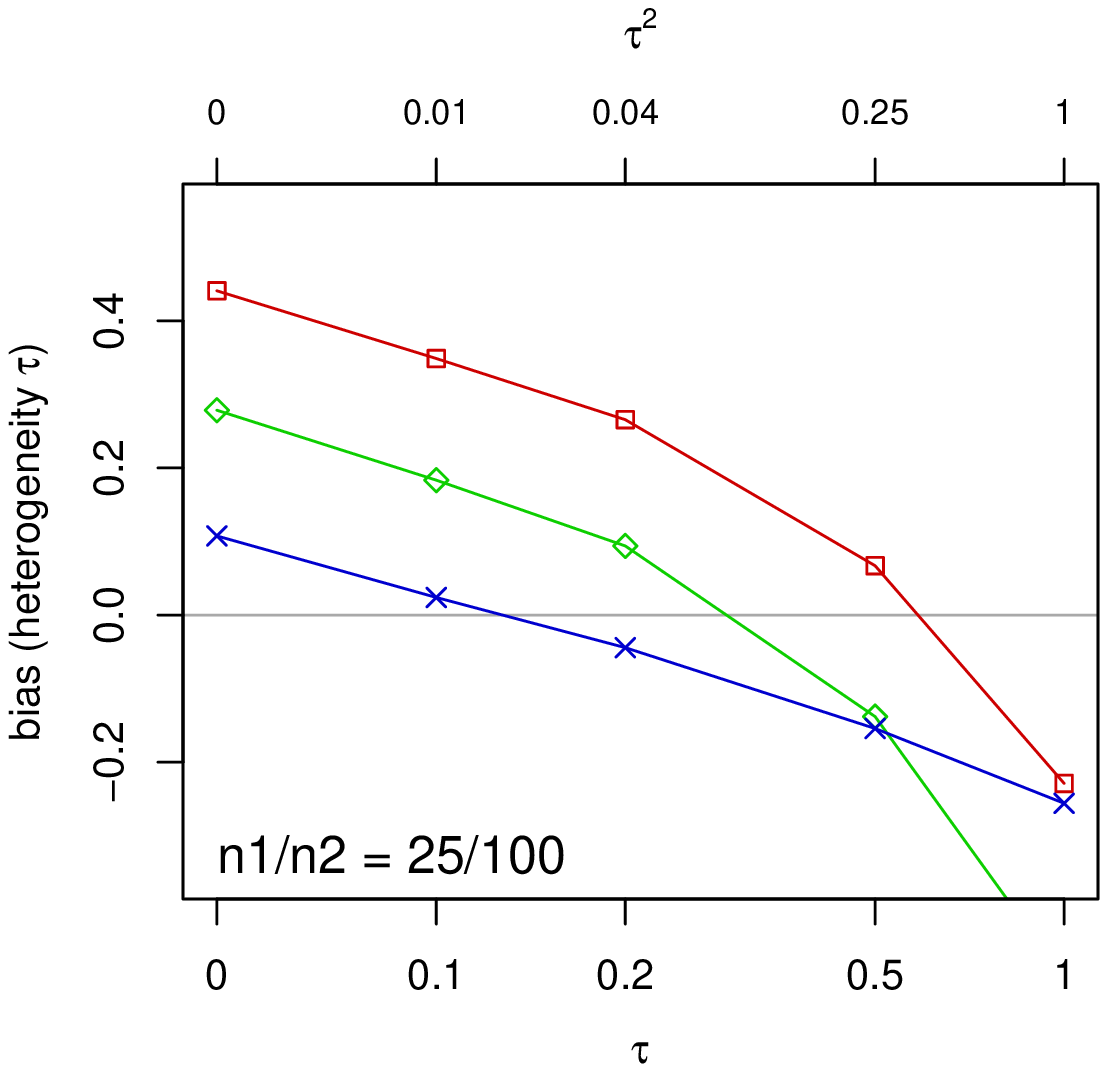} 
\includegraphics[width=0.3\linewidth]{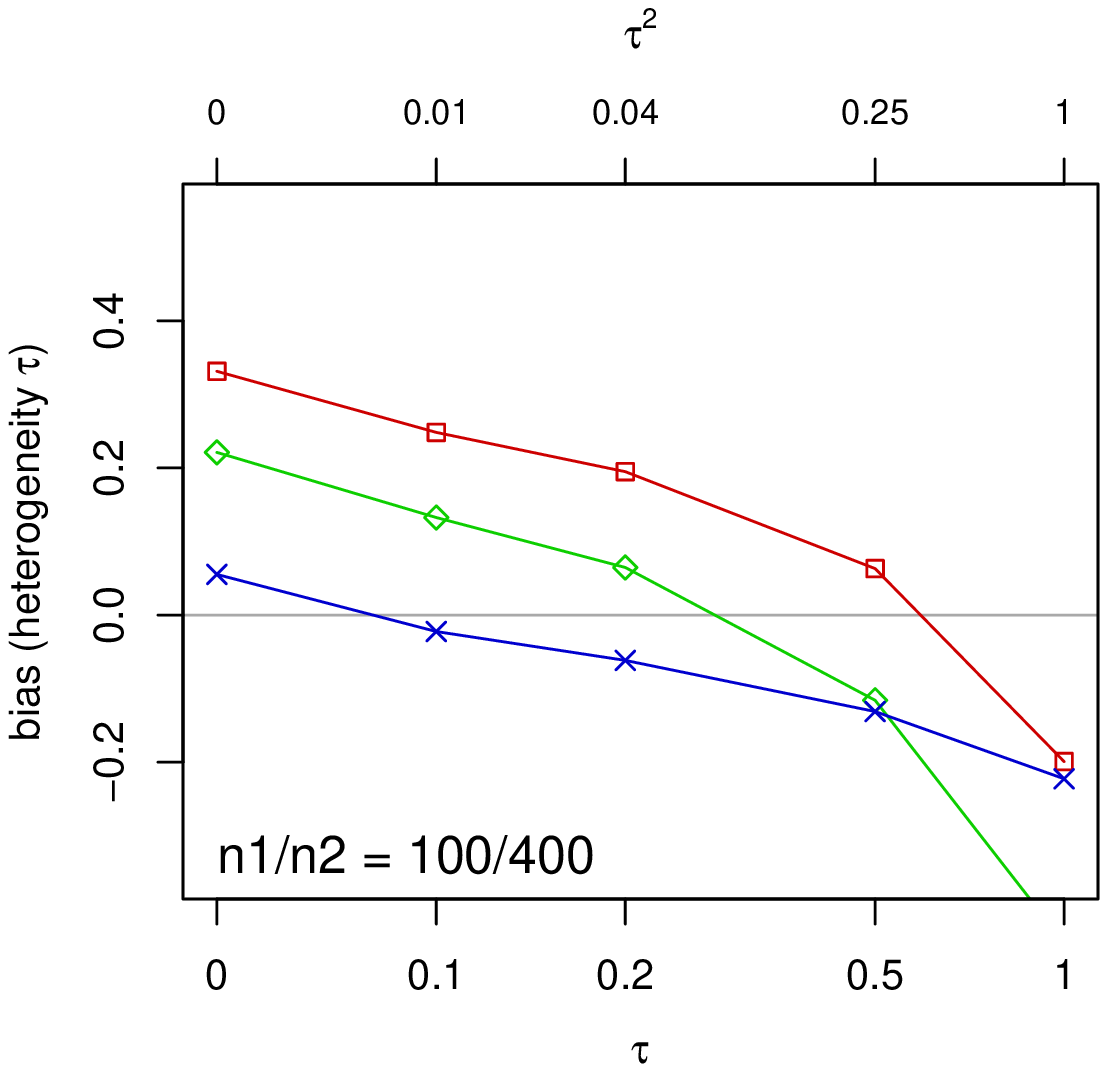} 
\includegraphics[width=0.3\linewidth]{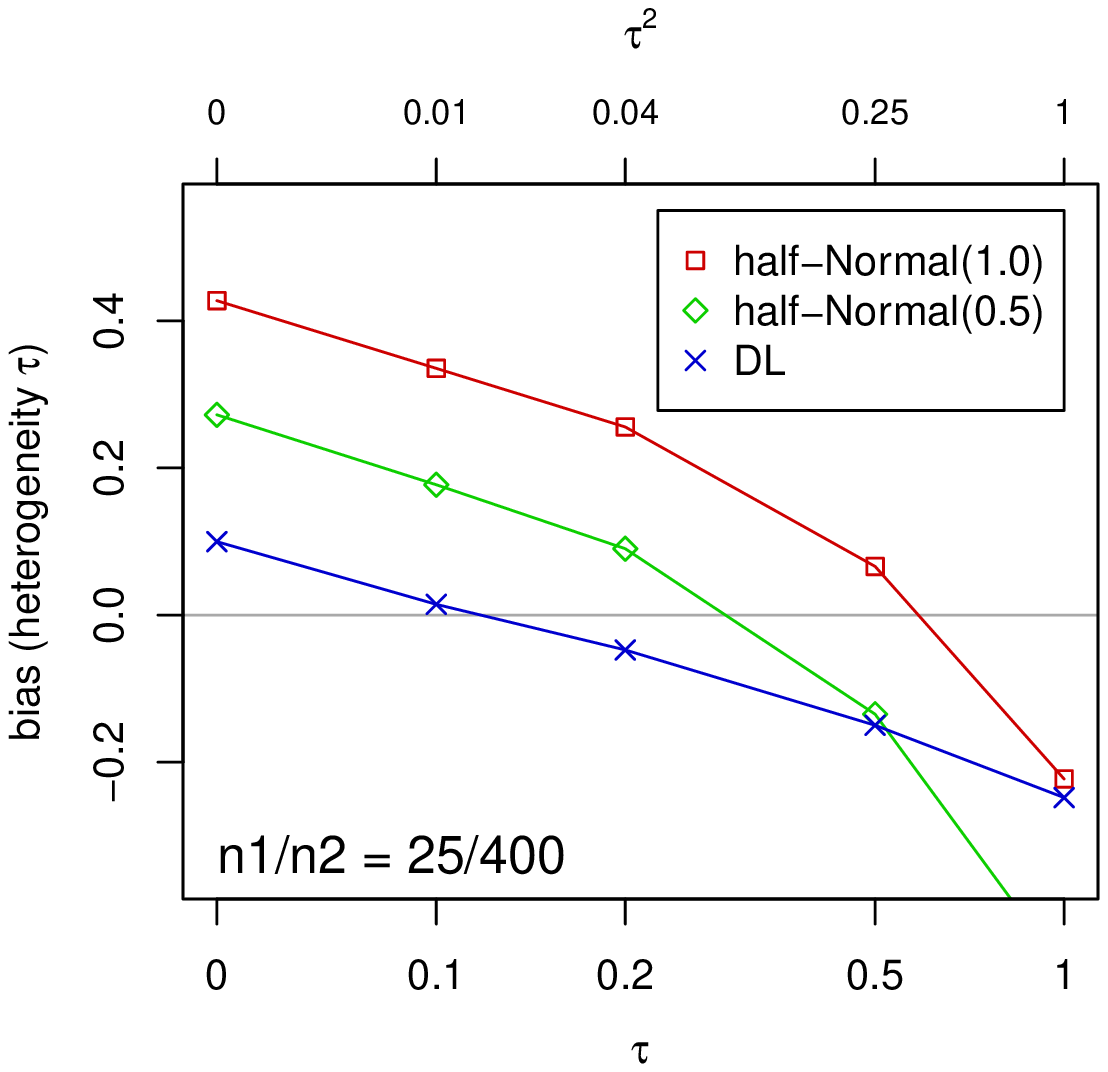}
\caption{Bias in estimating the between-study heterogeneity $\tau$ in the different simulation settings.} \label{fig:bias}
\end{center}
\end{figure}

  Figure~\ref{fig:bias} shows the bias of $\tau$~estimates. The DL
  estimator tends to overestimate $\tau$ if heterogeneity is small
  ($\tau=0, 0.1$). On the other hand, $\tau$ will be underestimated
  for substantial ($\tau=0.5$) and large ($\tau=1$) heterogeneity. The
  Bayesian estimators (posterior medians) show similar patterns, with
  the magnitude of bias depending on the prior. The overestimation
  under small heterogeneity is obviously more pronounced for the
  HN(1.0) than for the HN(0.5) prior, because the former favours larger values of $\tau$. On the other hand, underestimation
  of $\tau$ only occurs (and is fairly small) if heterogeneity is
  substantial to large. It should be noted that, in contrast to the
  frequentist methods, the Bayesian estimates for $\tau$ are less
  important, since the inference for $\mu$ takes into account the
  uncertainty of $\tau$ via the posterior distribution.

\subsection{Fraction of zero $\tau$ estimates}
  Table~\ref{tab:zeroes} shows that the DL estimates for $\tau$ are
  often zero even if heterogeneity is substantial $(\tau=0.5)$ or
  large $(\tau=1.0)$. This is a well-known problem, which, if not
  appropriately addressed, results in too optimistic inferences
  for~$\mu$.
  \begin{table}[ht]
    \begin{center}
      \caption{\label{tab:zeroes} Fractions (in~\%) of heterogeneity 
               estimates turning out as zero.}
      \begin{tabular}{r@{ / }lrrrrr}
        \hline \hline
        \multicolumn{2}{c}{ }& \multicolumn{5}{c}{true heterogeneity~$\tau$} \\
        $n_1$ & $n_2$ & 0.0 & 0.1 & 0.2 & 0.5 & 1.0 \\ 
        \hline
        25 & \phantom{0}25 & 68 & 67 & 62 & 47 & 29 \\ 
        100 & 100 & 68 & 63 & 52 & 29 & 15 \\ 
        400 & 400 & 68 & 53 & 34 & 16 & \phantom{0}8 \\[1.0ex]
         25 & 100 & 68 & 65 & 60 & 41 & 23 \\ 
        100 & 400 & 68 & 61 & 46 & 24 & 13 \\ 
         25 & 400 & 68 & 65 & 59 & 39 & 22 \\ 
        \hline \hline
      \end{tabular}
    \end{center}
  \end{table}

\subsection{Coverage probabilities and interval lengths for $\mu$}
\begin{figure}[ht]
    \begin{center}
      \includegraphics[width=0.3\linewidth]{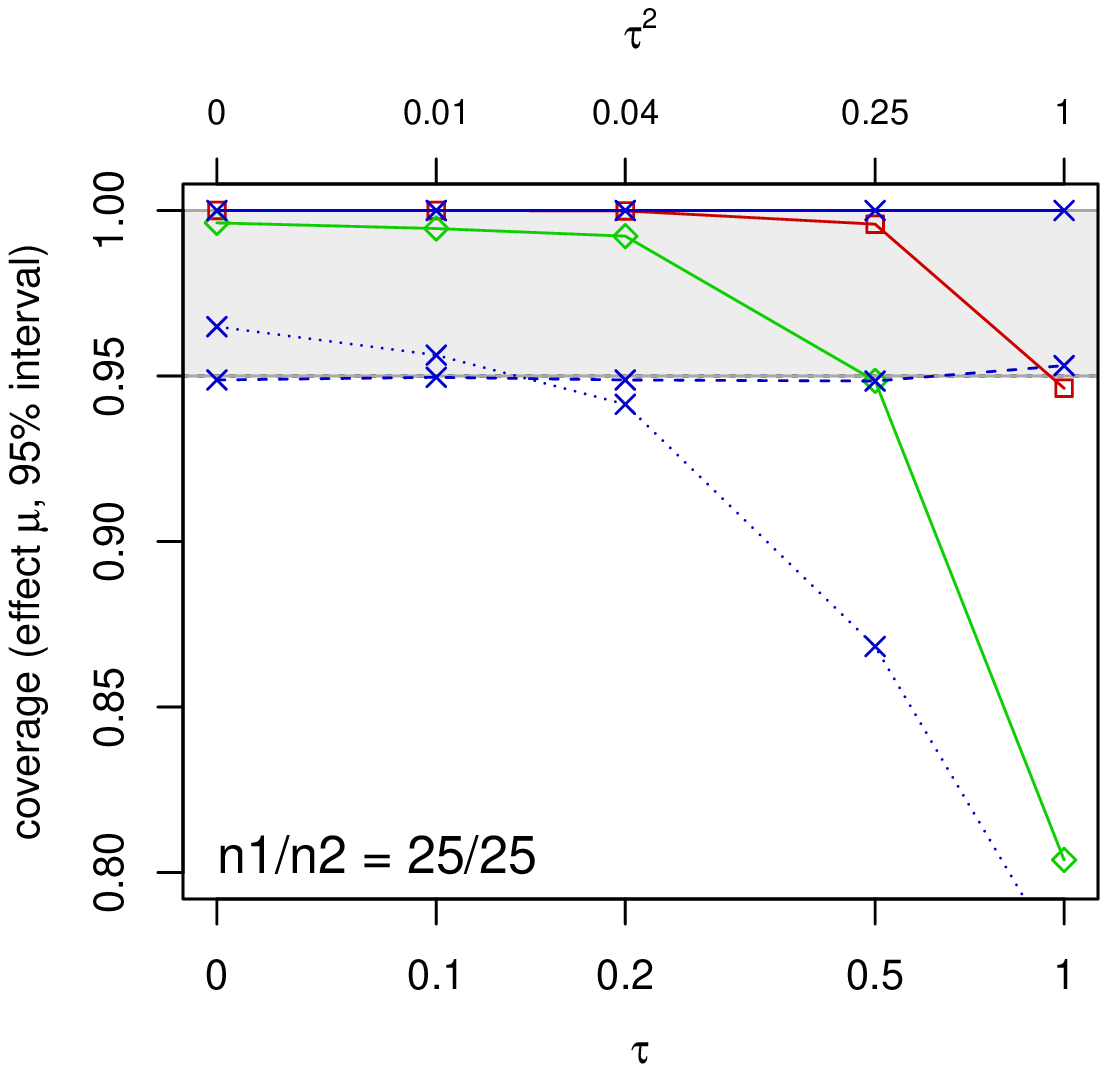} 
      \includegraphics[width=0.3\linewidth]{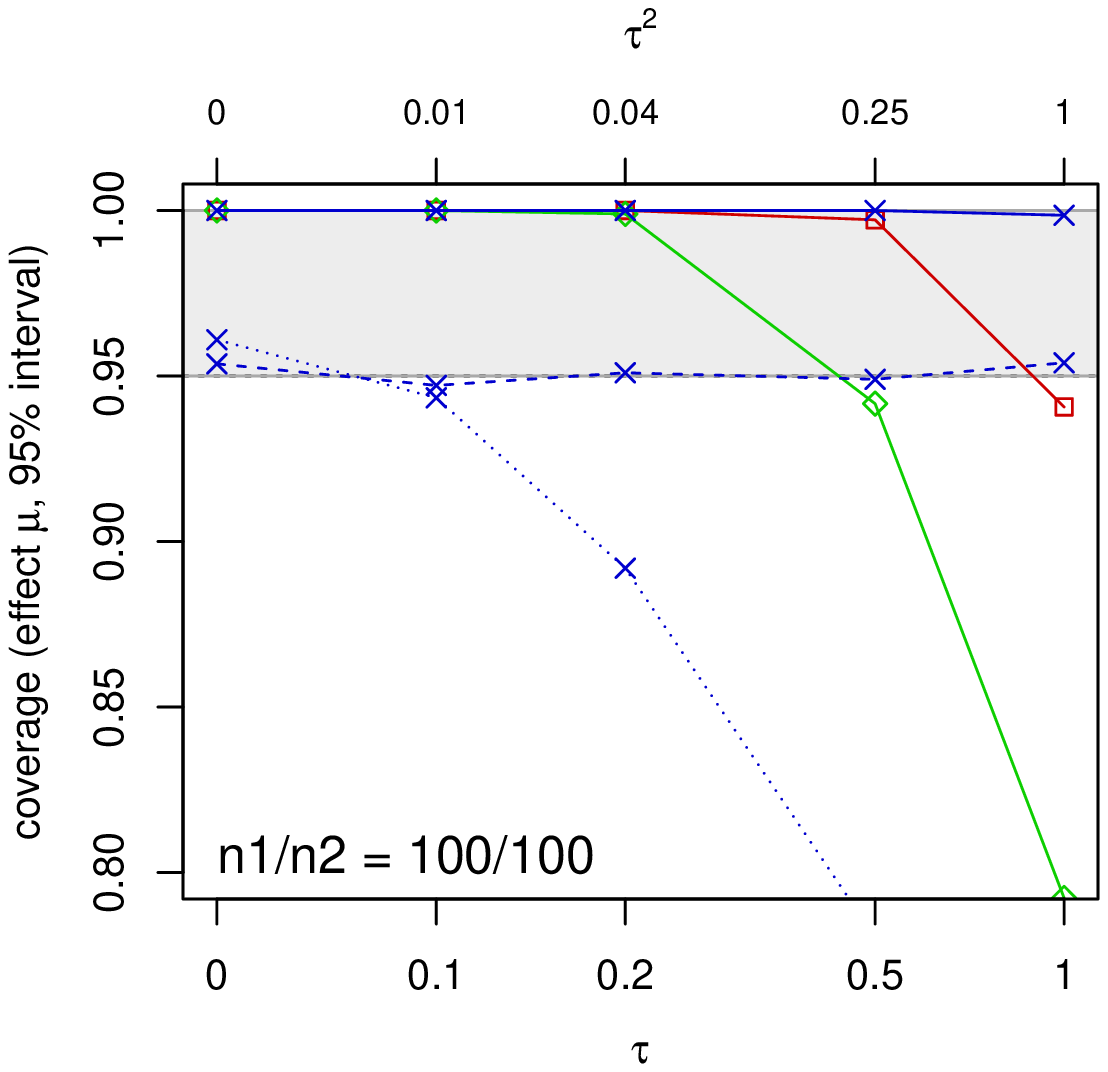} 
      \includegraphics[width=0.3\linewidth]{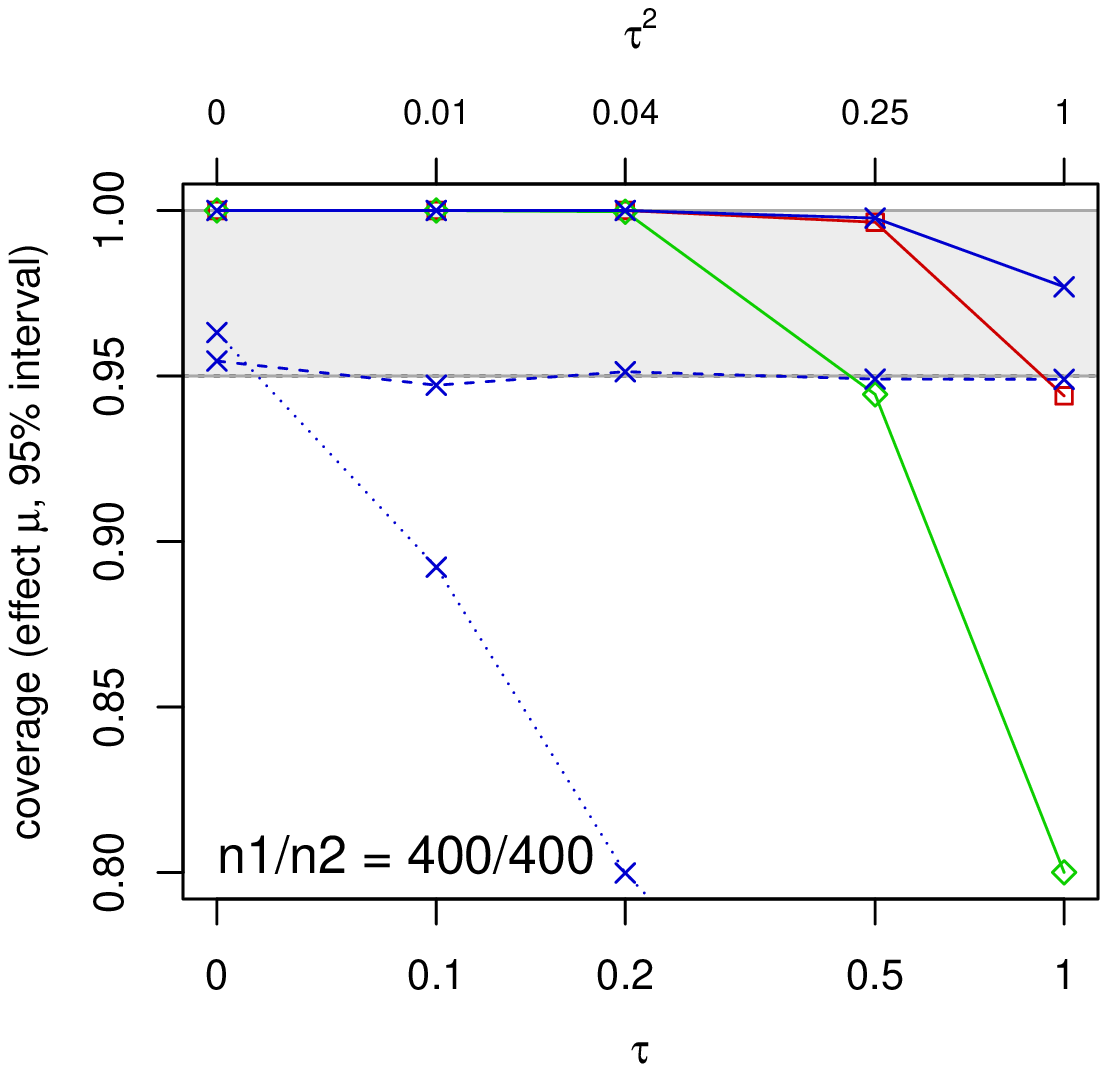} \\
      \includegraphics[width=0.3\linewidth]{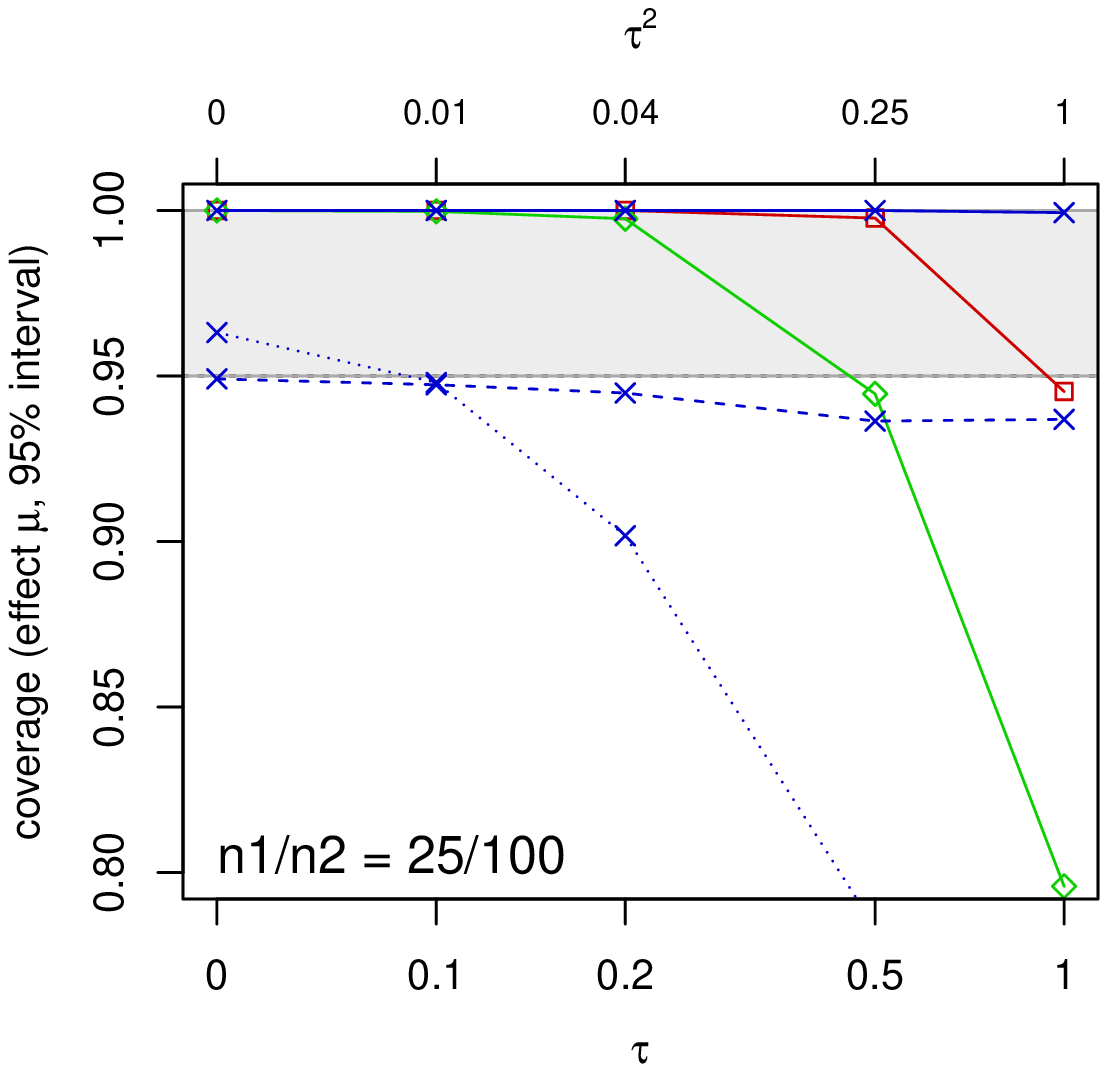} 
      \includegraphics[width=0.3\linewidth]{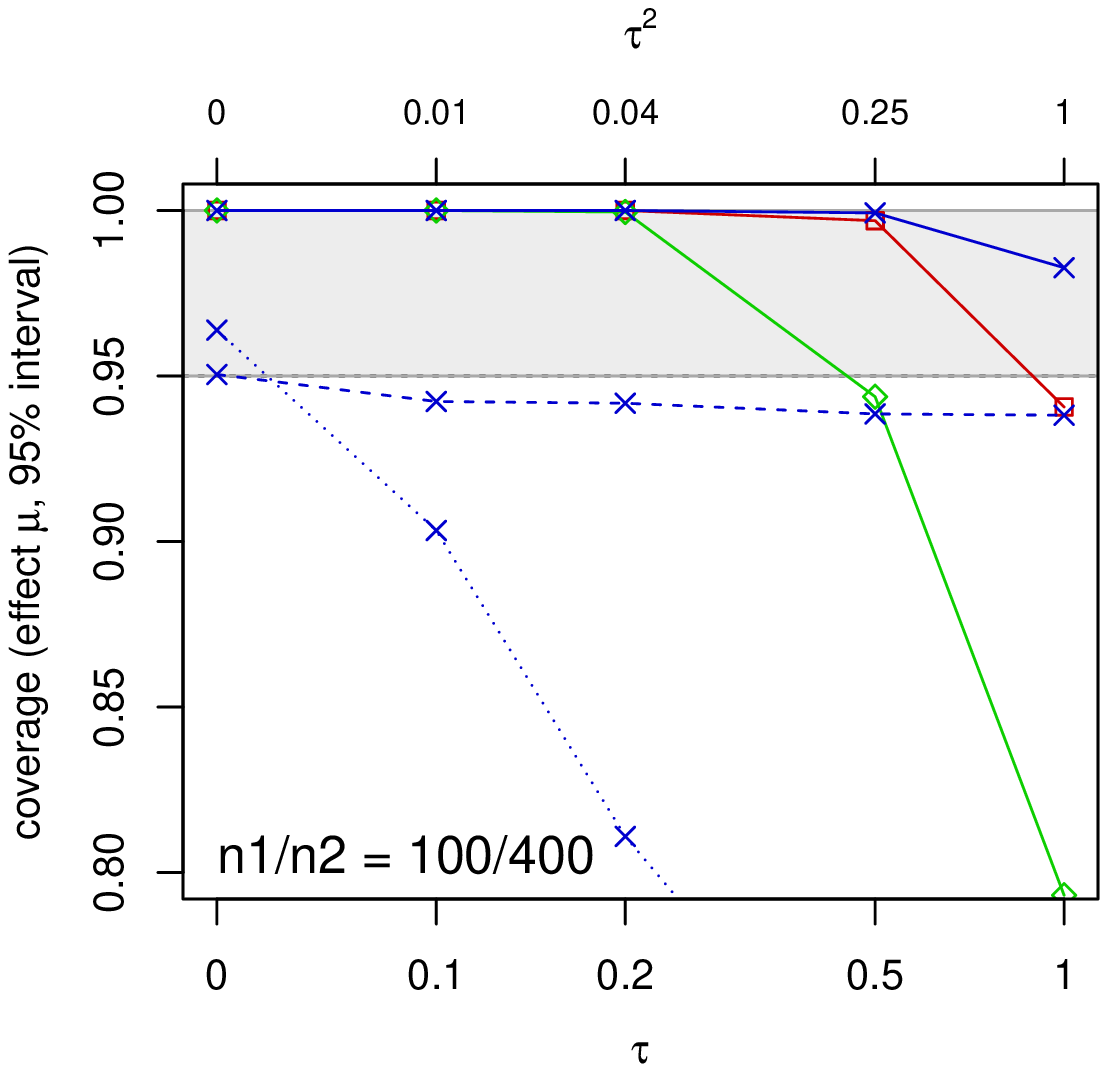} 
      \includegraphics[width=0.3\linewidth]{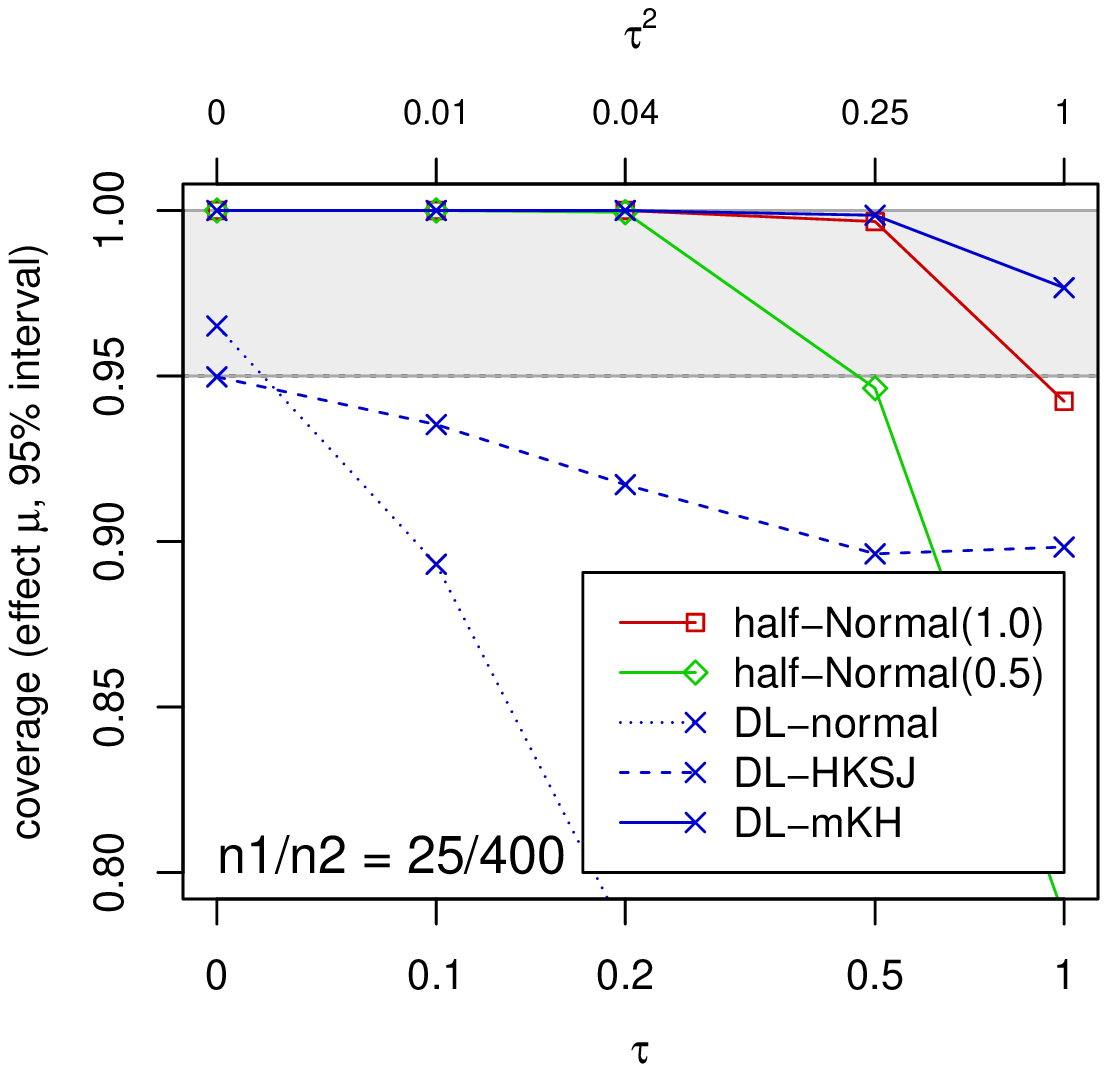}
      \caption{Coverage of interval estimators in the different simulation settings.} \label{fig:cicoverage}
    \end{center}
  \end{figure}

  As can be seen from Figure~\ref{fig:cicoverage}, if heterogeneity is
  small, all methods work well save for the normal approximation
  (DL-normal), for which the coverage can be below the nominal level
  even for small heterogeneity $(\tau=0.1)$.  The HKSJ method is known
  to work well for equally sized studies, but can be problematic for
  unequal study sizes and considerable heterogeneity. As can be seen,
  the modified method (mKH) resolves this problem. The Bayesian
  intervals show good coverage in the range of the prior, irrespective
  of the study sizes. For example, under the more optimistic HN(0.5)
  prior, coverage is reasonable for $\tau$ up to 0.5, but will drop
  considerably for larger values that are a-priori less likely.

\begin{figure}[ht]
    \begin{center}
      \includegraphics[width=0.3\linewidth]{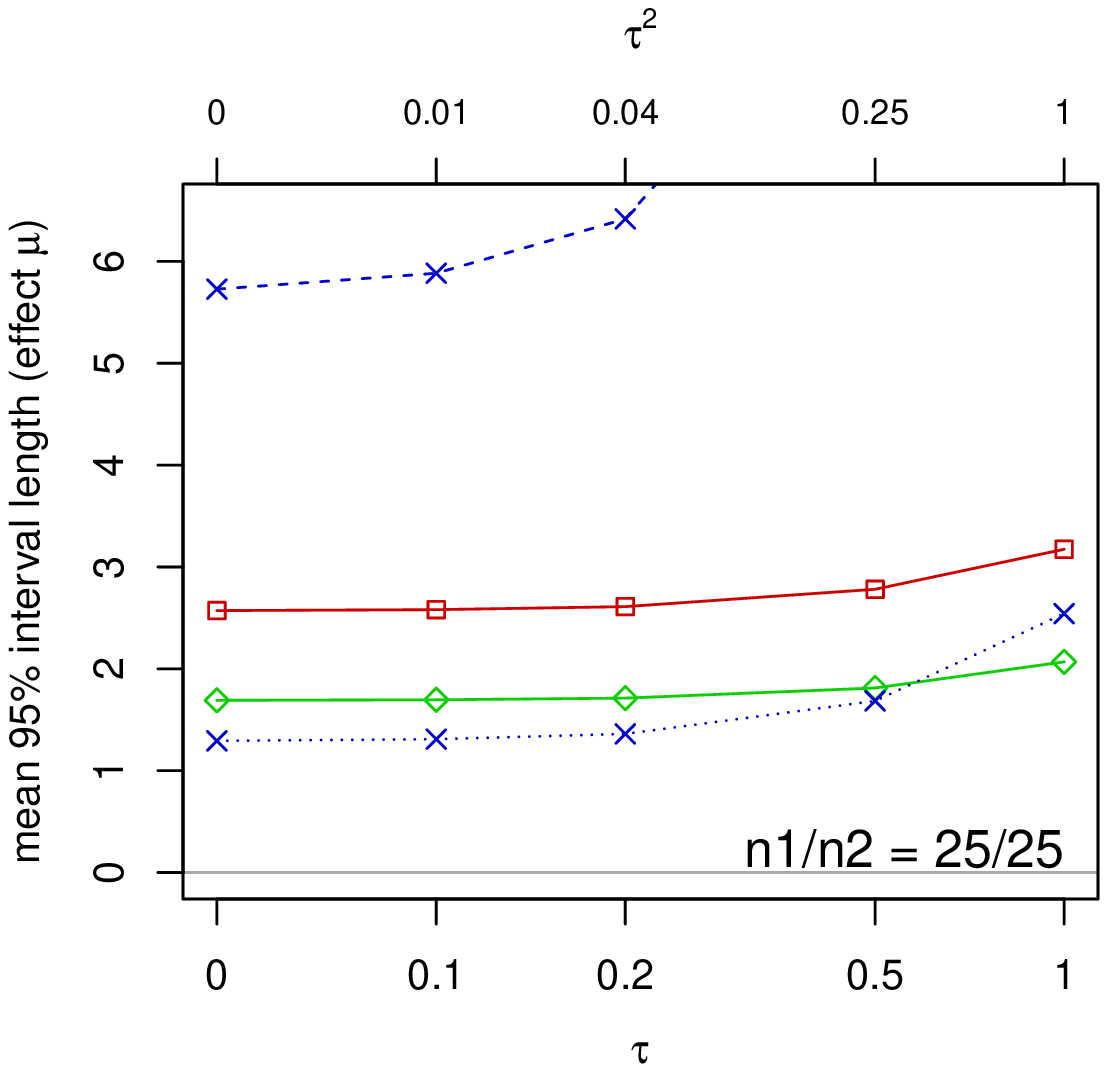} 
      \includegraphics[width=0.3\linewidth]{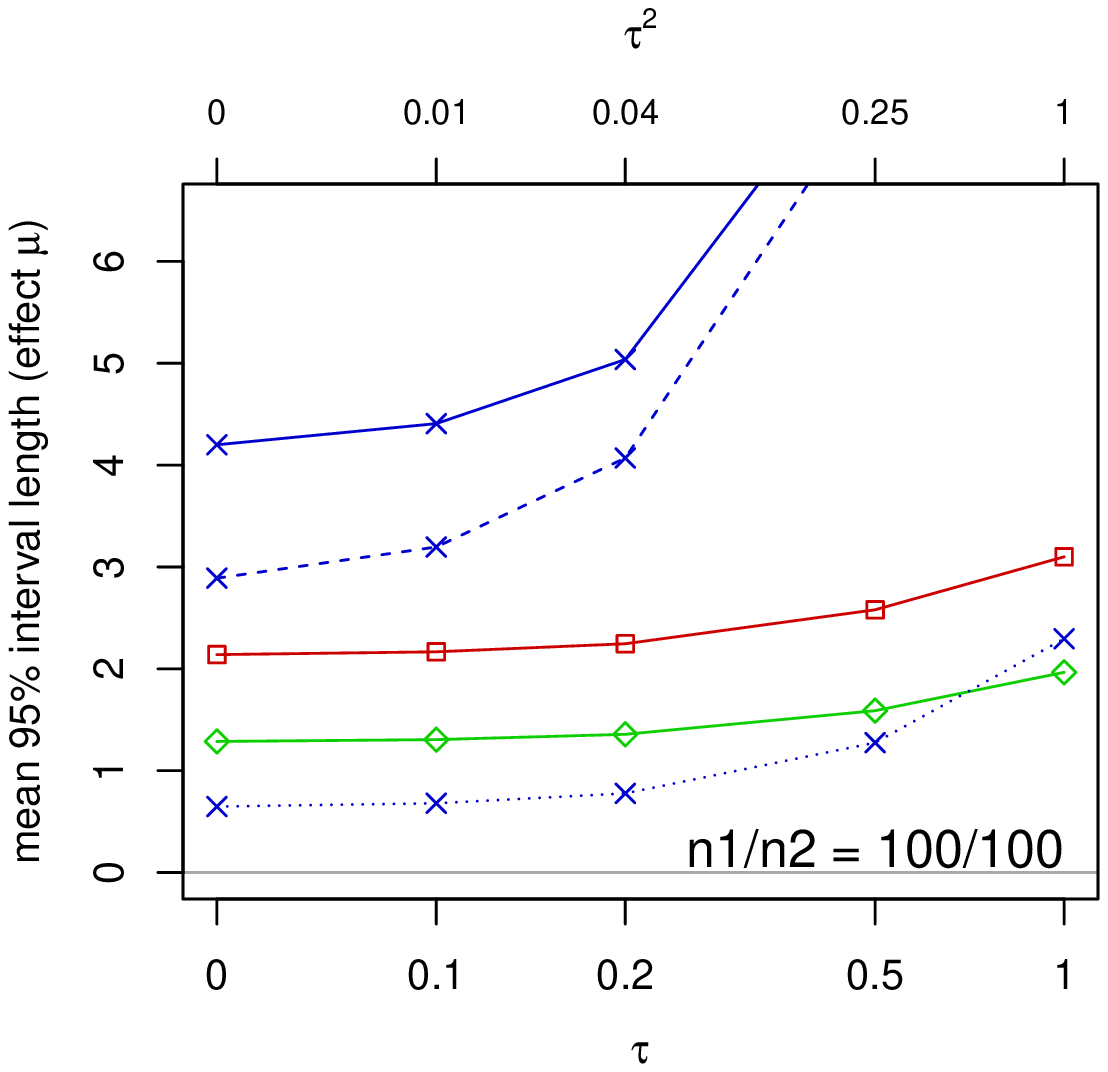} 
      \includegraphics[width=0.3\linewidth]{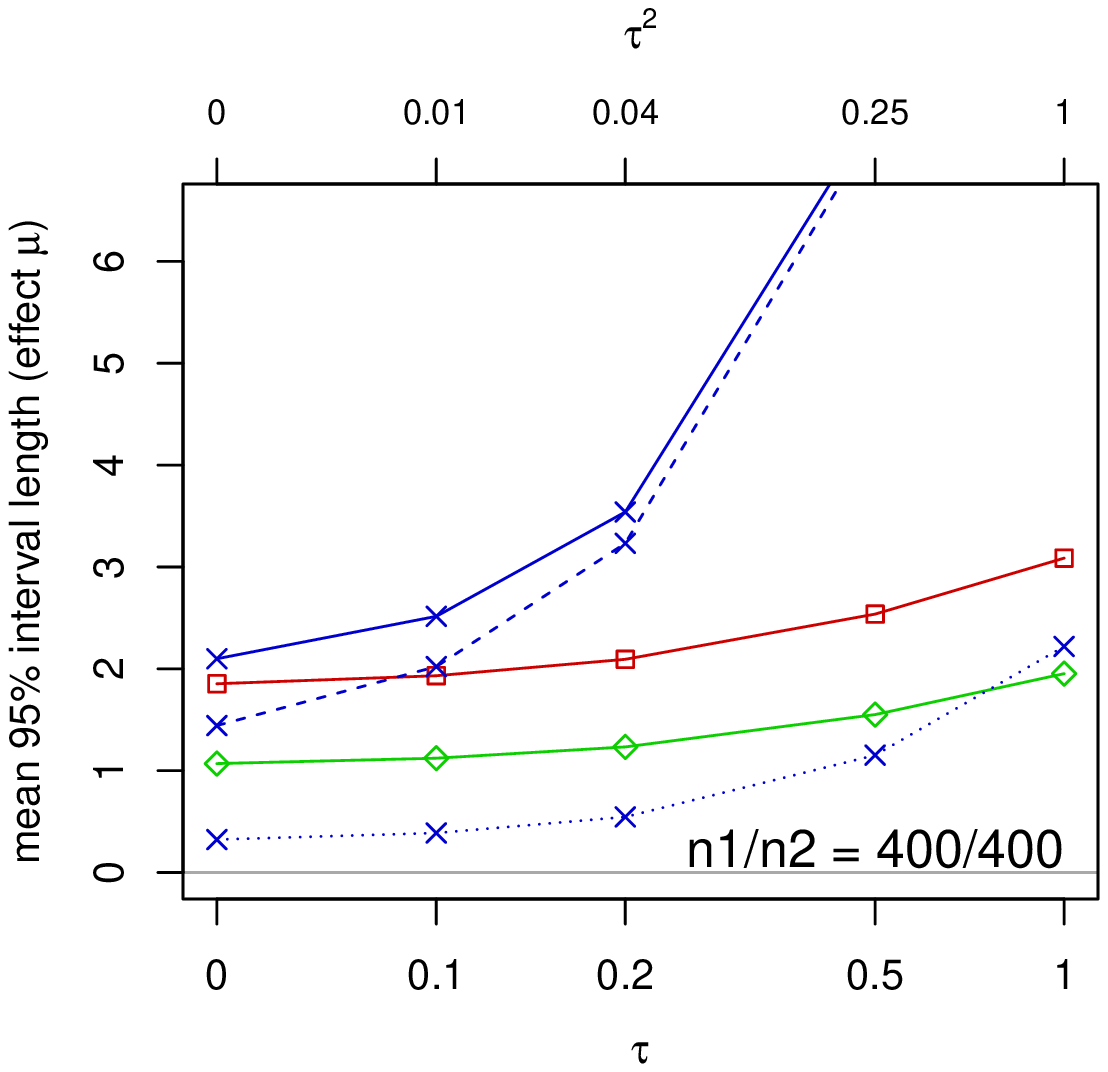} \\
      \includegraphics[width=0.3\linewidth]{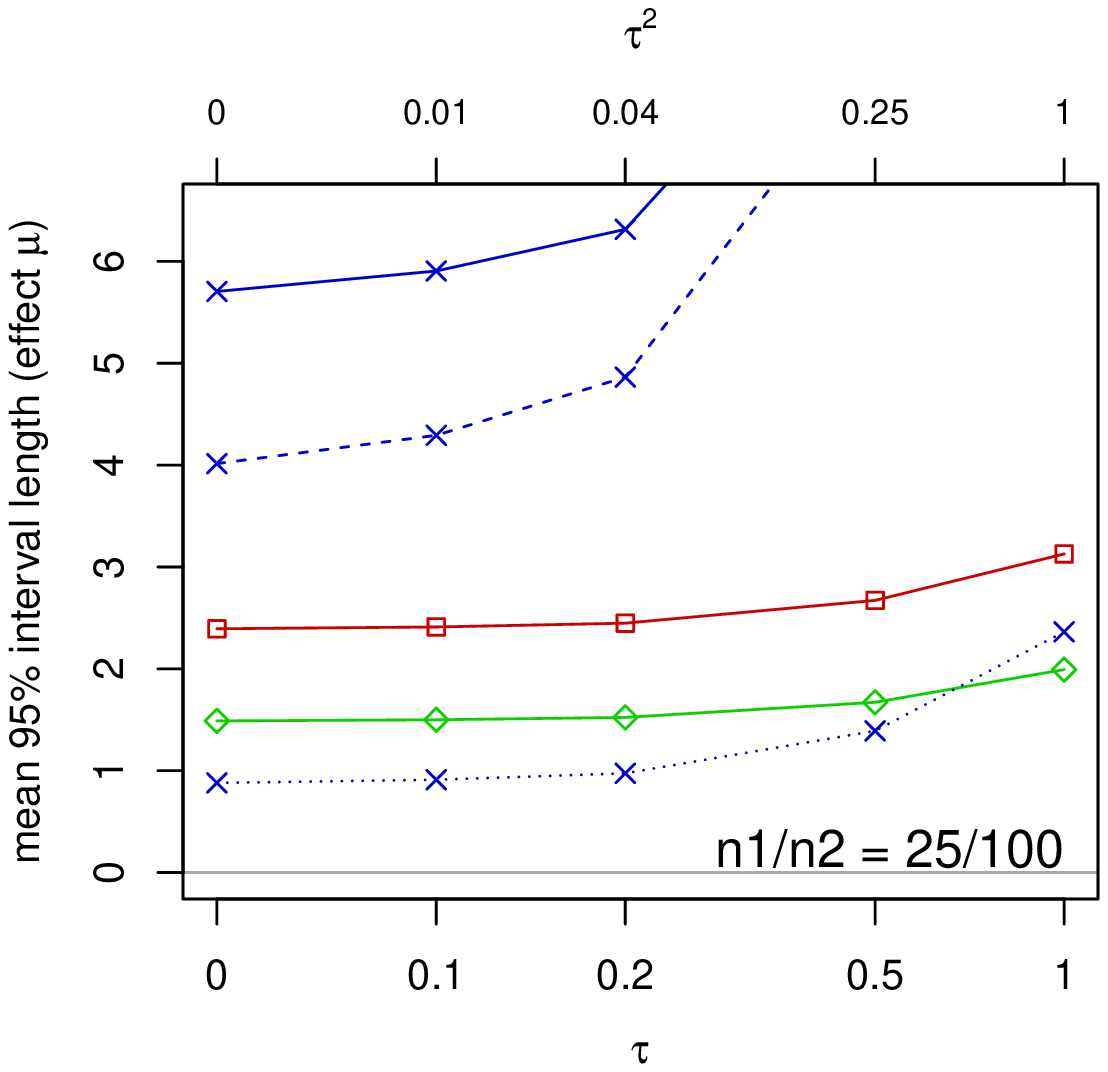} 
      \includegraphics[width=0.3\linewidth]{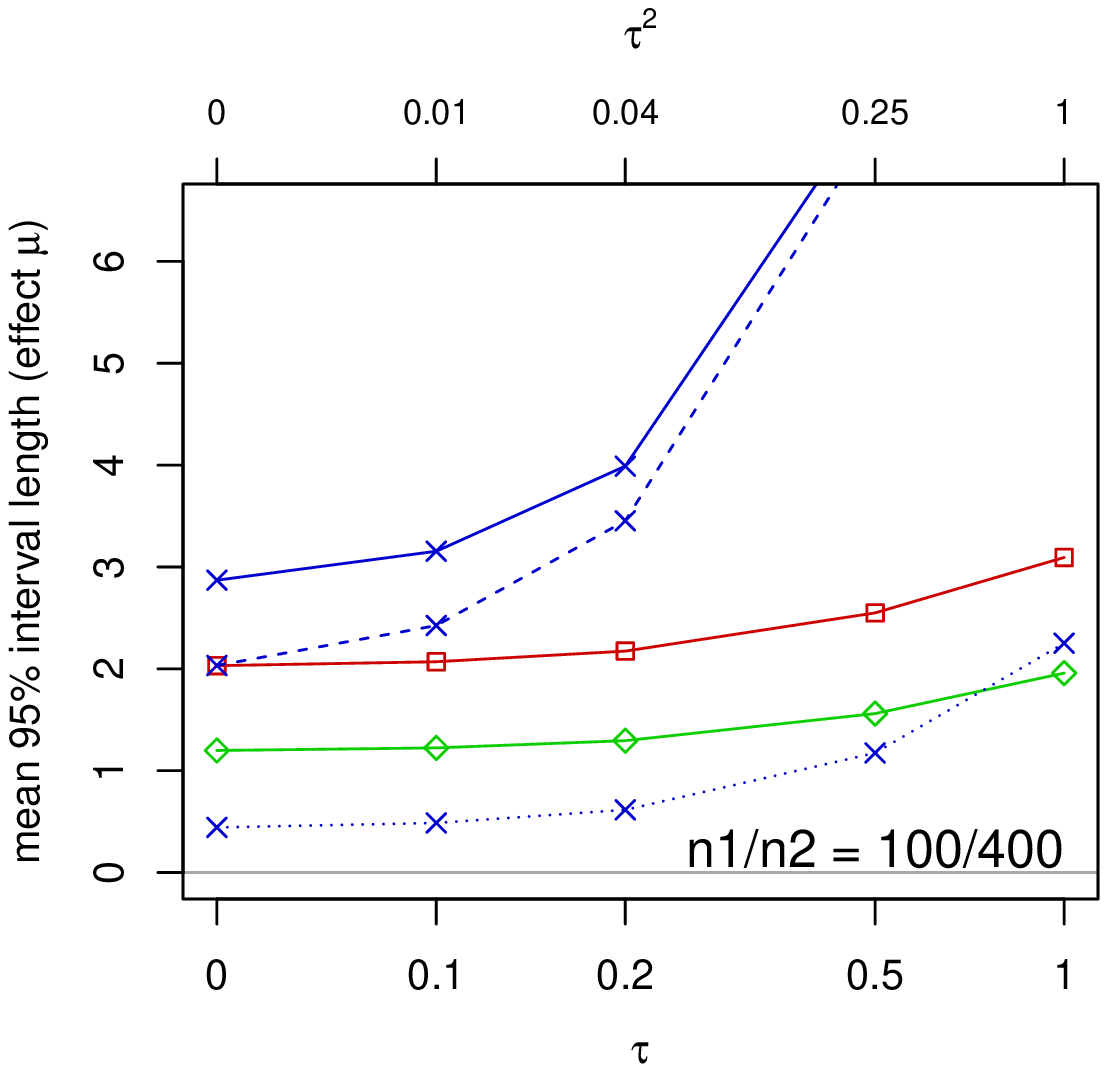} 
      \includegraphics[width=0.3\linewidth]{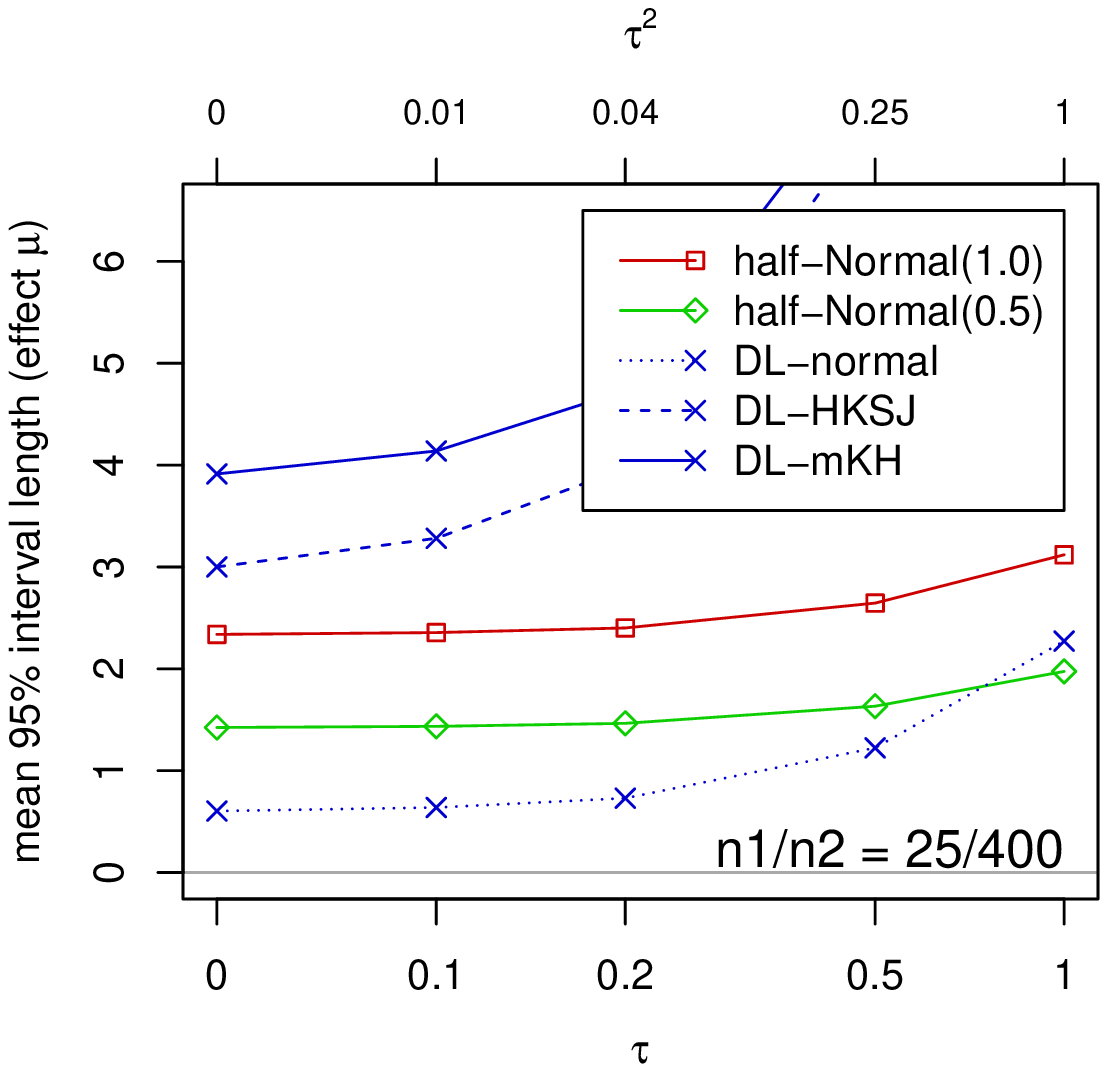}
      \caption{Mean lengths of confidence / credibility intervals in the different simulation settings.} \label{fig:cilength}
    \end{center}
  \end{figure}

  Coverage is obviously linked to interval length: higher coverage
  generally comes at the price of longer
  intervals. Figure~\ref{fig:cilength} shows that this price can be
  very high. The two frequentist methods with good coverage (HKSJ,
  mKH) exhibit exorbitantly long and implausible 95\%-intervals, for
  which practical relevance is unclear.  Interestingly, the Bayesian
  intervals are much shorter and provide a sensible compromise between
  the HSJK or mKH and the DL-normal intervals. These findings are
  consistent with results of the examples in
  Section~\ref{sec:examples}.

\section{Discussion} \label{sec:discussion}
  There is a need for random-effects meta-analyses with only two
  studies, in particular in rare diseases. To gain insights into the
  properties of various meta-analytic methods for two trials, in this
  special case we considered examples from the area of rare diseases
  and conducted an extensive simulation study.  The simulations
  allowed us to assess the coverage probabilities and mean lengths of
  the confidence and credibility intervals. The 
  examples led to further insights into the interpretability
  of results.

  We can summarize our findings as follows. The confidence intervals
  based on normal quantiles do not have the right coverage and cannot
  be recommended for use in the case of two studies. The HKSJ
  intervals provide good coverage if the standard errors of the
  treatment effects observed in the two studies are of similar
  size. In general, however, the HKSJ intervals are either so wide
  that they do not allow any conclusion, or are very narrow. The
  latter occurs rarely (if the two study estimates are very
  close,~(\ref{eqn:HKSJ})), but can lead to problematically narrow
  confidence intervals and unfavourable coverage. This can be fixed by
  the ad-hoc modification (mKH), which is in agreement with findings
  by \citet{RoeverKnappFriede2015}. The mKH method yields generally
  coverage probabilities in excess of the nominal level, but the
  intervals are generally so wide that they do not allow any
  meangingful conclusion. In this sense we agree with
  \cite{GonnnermannEtAl2015} that there is currently no solution for
  random-effects meta-analysis in the frequentist setting. However,
  Bayesian random-effects meta-analyses with a reasonable prior yield
  interpretable results in our examples and showed satisfying
  properties in the simulations. Therefore, the Bayesian intervals appear
  to be a reasonable compromise between the extremes of the confidence
  intervals based on normal quantiles that suffer from poor coverage
  and the $t$-distribution based intervals that tend to be so long
  that they are inconclusive.
  Use of a Bayesian approach of course entails the question of what
  constitutes sensible prior information in a given context. This may
  be argued on the basis of the endpoint in question, i.e., what is
  the plausible amount of heterogeneity expected e.g.\ among log-ORs,
  as in the motivating examples above. Otherwise the problem may be to
  determine what constitutes relevant external data, and how this
  information may be utilized to formulate a prior, as was done
  e.g.\ by \citet{TurnerEtAl2012} and \citet{RhodesEtAl2015}.

  Here we investigated several meta-analytic methods for two studies,
  with a focus on rare diseases. While a definite answer to this
  challenging problem is under dispute, the proposed Bayesian approach
  works well in our examples and simulation settings. The current
  frequentist methods have severe limitations, which may be addressed
  with future research. Until these limitations are resolved, we
  recommend to meta-analyze two heterogeneous studies in a Bayesian
  way using plausible priors.

\section*{Acknowledgements}
This research has received funding from the EU's 7th Framework
Programme for research, technological development and demonstration
under grant agreement number FP HEALTH 2013602144 with project
title (acronym) ``Innovative methodology for small populations
research'' (InSPiRe). \\

\noindent {\bf{Conflict of Interest}}
\noindent \textit{BN and SW are employees of
Novartis, the manufacturer of an Interleukin-2 receptor antagonist
used in the examples.}

{ 
  \bibliographystyle{bimj}
  \bibliography{literature}
}

\end{document}